\documentclass[11pt,letterpaper]{article}
\usepackage[utf8]{inputenc}
\usepackage{style}
\usepackage{kotex}
\usepackage{algorithm}
\usepackage{algpseudocode}
\usepackage{todonotes}
\usepackage{booktabs}
\usepackage{graphicx}
\usepackage{thm-restate}
\usepackage{amsmath}

\algnewcommand{\Input}{\Statex\textbf{Input:}}
\algnewcommand{\Output}{\Statex\textbf{Output:}}

\DeclarePairedDelimiter{\bra}{\langle}{\rvert}
\DeclarePairedDelimiter{\ket}{\lvert}{\rangle}
\DeclarePairedDelimiterX\innerp[2]{\langle}{\rangle}{#1\delimsize\vert\mathopen{}#2}
\DeclarePairedDelimiterX\braket[2]{\langle}{\rangle}{#1\delimsize\vert\mathopen{}#2}
\DeclarePairedDelimiterX\braketOP[3]{\langle}{\rangle}{#1\,\delimsize\vert\,\mathopen{}#2\,\delimsize\vert\,\mathopen{}#3}
\DeclarePairedDelimiterX\ketbra[2]{\lvert}{\rvert}{#1\delimsize\rangle\!\delimsize\langle#2}
\DeclarePairedDelimiterX\outerp[2]{\lvert}{\rvert}{#1\delimsize\rangle\!\delimsize\langle#2}
\DeclarePairedDelimiterX\projector[1]{\lvert}{\rvert}{#1\delimsize\rangle\!\delimsize\langle#1}

\title{Shadow bootstrapping:\\
Nearly minimal-setting pure-state tomography\\
with shallow Clifford measurements}

\author{
Gyungmin Cho\thanks{
NextQuantum Center, Department of Physics and Astronomy, and Institute of Applied Physics, Seoul National University, Seoul 08826, Korea. \texttt{km950501@snu.ac.kr}
}
\and
Dohun Kim\thanks{
NextQuantum Center, Department of Physics and Astronomy, and Institute of Applied Physics, Seoul National University, Seoul 08826, Korea. \texttt{dohunkim@snu.ac.kr}
}
}

\date{\today}

\begin{document}
\allowdisplaybreaks
\maketitle

\begin{abstract}\normalsize
     We present a pure-state tomography algorithm that uses a nearly minimal number of measurement settings while achieving optimal sample complexity with nonadaptive single-copy shallow Clifford measurements. For an arbitrary $n$-qubit pure state $\ket{\psi}\in \mathbb{C}^{d}$, where $d=2^n$, the algorithm uses $O(d/\epsilon)$ copies and $\widetilde{O}(1)$ measurement settings to output a state vector $|\hat{\psi}\rangle$ satisfying $1-|\langle\psi|\hat{\psi}\rangle|^2\le\epsilon$ with high probability. Each measurement is implemented by a Clifford circuit of depth $O(\log n)$. 
     Reusing each measurement setting allows the data to be stored compactly as per-basis outcome histograms. Given these histograms, classical post-processing runs in $\widetilde{O}(d)$ time. Our key technique, \textit{shadow bootstrapping}, uses the current estimate to reduce the variance of subsequent residual estimation, enabling progressively more accurate reconstruction with only a constant number of new measurement settings per stage. 
\end{abstract}

\clearpage
\tableofcontents
\clearpage


\section{Introduction}\label{sec:introduction}

Quantum state tomography (QST) aims to obtain a classical description of an unknown quantum state. A major focus of QST research has been to minimize the number of copies required for accurate reconstruction. For an $n$-qubit pure state $\ket{\psi}$ in dimension $d=2^n$, the optimal sample complexity is $\Theta(d/\epsilon)$ for infidelity $1-|\langle\hat\psi|\psi\rangle|^2\le\epsilon$ and constant failure probability. This scaling can already be achieved using independent single-copy measurements~\cite{haah2016sample, odonnell2016tomography, scharnhorst2025optimal, haah2023query,guctua2020fast}.

Pure-state tomography also provides a way to learn mixed states. Recent reductions based on random purification channels~\cite{tang2025conjugate} produce copies of the same purification of a mixed state. Learning this purification yields a classical description from which an estimate of the original mixed state can be recovered~\cite{pelecanos2025mixed}. The collective operations across copies are confined to the purification step; the remaining reconstruction uses pure-state tomography.

Another important resource for QST is the number of distinct measurement settings. Each setting specifies a unitary circuit followed by computational-basis readout, and the same circuit can be applied repeatedly to fresh copies.
Introducing a new setting can require circuit compilation and changes to the control program~\cite{flammia2012quantum}. Even with optimal sample complexity and shallow circuits, using many distinct measurement settings can add substantial experimental overhead.
This motivates reusing a smaller collection of measurement settings~\cite{chen2025optimizing,zhou2023performance, helsen2023thrifty,korhonen2025practical}.

Reusing the same settings also allows measurement outcomes to be stored more compactly.
A raw outcome list records every observed bit string $b\in\{0,1\}^n$ separately, along with the corresponding measurement basis $U$. Repeated measurements in each basis can instead be summarized by a histogram, with observed bit strings as keys and their occurrence counts as values.
Each histogram contains at most $d$ entries, and each count uses only logarithmically many bits in the number of shots. For example, the outcomes of $O(d/\epsilon)$ shots in a single basis can be stored in $\widetilde O(d)$ memory. These histograms can be accumulated directly during the experiment without storing a raw outcome list.
Working directly with this smaller input can reduce the cost of classical reconstruction as well as its memory requirements.

Learning from few measurement bases has been studied
in several settings. A constant number of bases can suffice for pure-state identification and reconstruction from exact outcome distributions~\cite{carmeli2015many,goyeneche2015five, carmeli2016stable,voroninski2013quantum, kech2017constrained,baldwin2016strictly}.
Some of these works bound the reconstruction error when the outcome probabilities are known only approximately, but they do not provide explicit sample complexity bounds in terms of dimension and target infidelity. More recently, finite-sample protocols using randomly chosen bases have reduced the dimension dependence of the number of measurement settings, but this number still grows polynomially with inverse accuracy~\cite{brandao2020fast,cho2026quantum}.

Here, we aim to improve this accuracy dependence by taking more shots per measurement basis, while keeping both the total number of samples and the classical runtime small. Our analysis uses shadow estimators~\cite{huang2020predicting} constructed from randomized measurements~\cite{elben2023randomized}, with multiple shots taken in each sampled basis. For these estimators, taking more shots reduces the noise in the empirical outcome bit string frequencies.
However, the estimator can still fluctuate with the random choice of bases, even if the outcome probabilities in each sampled basis are known exactly.
Establishing accurate reconstruction with few settings therefore requires controlling both the finite-shot noise and these fluctuations.

On the computational side, compact histogram input does not automatically lead to efficient classical post-processing. Explicitly forming a shadow estimator as a dense $d\times d$ matrix would already require $d^2$ entries, losing the computational benefit of the compact input.
To achieve nearly linear reconstruction time, we must exploit the measurement structure to recover a length-$d$ state vector directly from the histograms without forming such a matrix.

In this work, we give a sample-optimal pure-state tomography algorithm using shallow measurements, a nearly minimal number of measurement settings, and nearly linear classical post-processing time from outcome histograms.
To our knowledge, this is the first algorithm to achieve these guarantees simultaneously.

\begin{theorem}[Informal, see Theorem~\ref{thm:full-confidence}]\label{thm:main}
    For any $n$-qubit pure state $\lvert\psi\rangle\in\mathbb C^d$, where $d=2^n$, and any $\epsilon\in(0,1)$, there is a tomography algorithm using nonadaptive single-copy Clifford measurements of depth $O(\log n)$ that outputs a unit vector $\lvert\hat\psi\rangle$ satisfying $1-|\langle\hat\psi|\psi\rangle|^2\le\epsilon$ with probability at least $1-\delta$. Its measurement and computational resources satisfy
    \[
        N=\widetilde O(1),
        \qquad
        T=O(d\log(1/\delta)/\epsilon),
        \qquad
        \mathrm{Time}_{\mathrm{hist}}=\widetilde O(d),
    \]
    where $N$ is the number of sampled measurement settings, $T$ is the total number of copies, and $\mathrm{Time}_{\mathrm{hist}}$ is the classical post-processing time given per-setting outcome histograms. Here, $\widetilde O(\cdot)$ suppresses polylogarithmic factors in $d$, $1/\epsilon$, and $1/\delta$.
\end{theorem}

\subsection{Related Work}\label{sec:related-work}
\begin{table}[H]
    \centering
    \resizebox{\textwidth}{!}{%
    \begin{tabular}{ccccc}
        \toprule
        Reference
        & Total copies ($T$)
        & Settings ($N$)
        & Circuit depth
        & Classical time
        \\
        \midrule
        
        \cite{brandao2020fast}
        & $\widetilde O(d/\epsilon^2),\ \delta=1/3$
        & $\widetilde O(1/\epsilon)$
        & $O(\log d)$
        & $\widetilde O(d^2/\epsilon^{5/2})$
        \\[5pt]

        \cite{haah2023query}
        & $O(d/\epsilon),\ \delta=d^{-100}$
        & $O(d/\epsilon)$
        & $\operatorname{poly}(\log d,\log(1/\epsilon))$
        & $\operatorname{poly}(d,1/\epsilon)$
        \\[5pt]

        \cite{pelecanos2025mixed}
        & $O\!\left((d+\log(1/\delta))/\epsilon\right)$
        & $O\!\left((d+\log(1/\delta))/\epsilon\right)$
        & $\operatorname{poly}(d)$
        & $\operatorname{poly}(d,1/\epsilon)$
        \\[5pt]

        \cite{cho2026quantum}
        & $\widetilde O(d\log(1/\delta)/\epsilon)$
        & $\widetilde O(1/\epsilon)$
        & $O(\log\log d)$
        & $\operatorname{poly}(d)$
        \\[5pt]

        \cite{grewal2026nearly}
        & $\widetilde O(d\log(1/\delta)/\epsilon)$
        & $\widetilde O(d/\epsilon)$
        & $O(1)$
        & $\widetilde O(d/\epsilon)$
        \\[5pt]

        \textbf{This work}
        & $O\!\left(d\log(1/\delta)/\epsilon\right)$
        & $\widetilde O(1)$
        & $O(\log\log d)$
        & $\widetilde O(d)^*$
        \\
        \bottomrule
    \end{tabular}%
    }

    \caption{
        Resource upper bounds for $n$-qubit pure-state tomography
        with infidelity $\epsilon$ and failure probability
        at most $\delta$, where $d=2^n$.
        Here $\widetilde O$ suppresses polylogarithmic
        factors in $d$, $1/\epsilon$, and $1/\delta$. Settings denote upper bounds on the number of distinct
        measurement bases. The classical runtime marked with an asterisk (*) assumes per-basis outcome histograms as input.
    }
    \label{tab:resource-comparison}
\end{table}

\paragraph{Tomography from few measurement bases.}
Compressed-sensing approaches exploit the low-rank structure of density matrices to reconstruct quantum states from an incomplete set of observable expectation values~\cite{ gross2010quantum,liu2011universal,flammia2012quantum,kueng2017low}.
For Pauli observables, $\widetilde O(rd)$ expectation values suffice to recover a rank-$r$ state~\cite{flammia2012quantum}.
The number of measurement bases is a related but different resource, since each basis gives access to a full distribution over $d$ outcomes.
A complementary line of work focused directly on the number of measurement bases, showing that a constant number suffices to distinguish pure states~\cite{carmeli2015many,goyeneche2015five,carmeli2016stable, voroninski2013quantum,kech2017constrained,baldwin2016strictly}.
These works do not, however, derive explicit bounds on the number of copies required as a function of $d$ and the target infidelity $\epsilon$.

The Hamiltonian-updates approach of~\cite{brandao2020fast} reduces both the number of measurement settings and classical computational costs, with rigorous finite-sample guarantees. For pure-state infidelity, the bounds summarized in Table~\ref{tab:resource-comparison} require $\widetilde O(1/\epsilon)$ settings and $\widetilde O(d/\epsilon^2)$ samples.
The number of settings thus depends only polylogarithmically on dimension, but the required number of samples has suboptimal dependence on $\epsilon$, and the classical runtime bound is superlinear in $d$. Their analyzed global Clifford and unitary $4$-design implementations also do not establish an $O(\log n)$ measurement-depth bound.

More recently, \cite{cho2026quantum} obtained few-setting tomography using Clifford measurements of depth $O(\log n)$.
Their pure-state specialization uses $\widetilde O(d/\epsilon)$ samples and $\widetilde O(1/\epsilon)$ settings.
The sample complexity is nearly optimal, but the number of settings still depends polynomially on inverse accuracy, and a nearly linear runtime guarantee for the full reconstruction is not established.
Our algorithm reduces the accuracy dependence of the number of settings to logarithmic while achieving optimal sample complexity and nearly linear reconstruction time from histograms.

\paragraph{Sample-optimal pure-state tomography.}
Independent single-copy measurements suffice for sample-optimal pure-state tomography~\cite{haah2023query,pelecanos2025mixed,guctua2020fast}. They average estimators of the form $(d+1)U^\dagger|b\rangle\langle b|U-I$ and recover a pure-state estimate from a top eigenvector of the resulting matrix. 
\cite{haah2023query} transfer high matrix-moment bounds from Haar-random measurements to efficiently implementable approximate designs and apply Markov's inequality. This gives $O(d/\epsilon)$ copy complexity with failure probability at most $d^{-100}$. \cite{pelecanos2025mixed} bound the moments of scalar quadratic forms and combine Bernstein's inequality with a covering-net argument. Their analysis achieves $O((d+\log(1/\delta))/\epsilon)$ copies, separating the confidence overhead from the dimension dependence. Both protocols independently sample a fresh measurement basis per copy, so their upper bounds on the number of sampled measurement bases have a linear dependence on $d$.
For each measurement,~\cite{haah2023query} use $\operatorname{poly}(n)$ gates, whereas~\cite{pelecanos2025mixed} use $\operatorname{poly}(d)$ gates. 
Neither construction establishes an $O(\log n)$ measurement-depth bound. In classical post-processing, both reconstruct the state from a top eigenvector of a $d\times d$ empirical matrix, with poly($d$)-time guarantees but no nearly linear reconstruction bound $\widetilde{O}(d)$.

\paragraph{Nearly time-optimal pure-state tomography.}
\cite{grewal2026nearly} gave an algorithm using nonadaptive Pauli measurements with sample complexity $\widetilde O(d/\epsilon)$. With a raw list of measurement outcomes as input, their algorithm runs in time $\widetilde O(d/\epsilon)$ nearly linear in the list length. At each level of their reconstruction tree, they use Frobenius distance estimation to determine the coefficients for combining the child estimates. Their distance estimator balances the number of sampled Pauli observables against the repetitions per observable across multiple scales to achieve nearly optimal sample complexity. 
\cite{hansen2026quantum} extended this line of work to recovering the leading eigenvector of a mixed state using nonadaptive Pauli measurements, achieving sample complexity and runtime nearly linear in $d$ for fixed accuracy and spectral gap. However, for pure-state tomography, both works give an upper bound of $\widetilde O(d/\epsilon)$ on the number of distinct measurement bases. Using shallow Clifford measurements, our algorithm requires only $\widetilde O(1)$ measurement settings while achieving the optimal $O(d/\epsilon)$ sample complexity.

\paragraph{Multi-shot shadow estimation and common randomized measurements.}
Multi-shot shadow estimation reuses each randomized measurement circuit for multiple shots and analyzes the resulting correlations and variance~\cite{helsen2023thrifty,zhou2023performance}. Repeated shots suppress sampling noise, but fluctuations due to the random choice of measurement basis remain.
Common randomized measurements (CRM)~\cite{vermersch2024enhanced} reduce these fluctuations by comparing experimental measurements with predictions from a reference state represented, for example, as a matrix product state (MPS) or a full classical description.
Using the same random bases correlates the measured and predicted contributions, reducing variance when the reference is sufficiently close to the target state.
\cite{vermersch2024enhanced} bound the variance of estimators for linear and multicopy observables in terms of the reference accuracy, and numerically demonstrate variance reduction when a reference state close to $\rho$ is available in advance.
In contrast, our algorithm progressively improves an estimate of an arbitrary unknown pure state through successive stages, using only tomography data and requiring no prior classical approximation of the target state.

\paragraph{Small measurement ensembles for classical shadows.}
Previous studies construct small collections of measurement bases for classical shadow estimation~\cite{zhang2024minimal, yang2026optimal, wang2024classical}. Our focus is how many bases are sampled and used in an experiment to achieve a target accuracy while maintaining optimal sample complexity. A small measurement collection alone does not guarantee this finite-sample efficiency. Conversely, our protocol samples and reuses only a few bases from a much larger Clifford ensemble, without requiring the ensemble itself to be small.

\section{Technical Overview}
\label{sec:technical-overview}

We first explain how shadow bootstrapping halves the target infidelity at each stage using a constant number of new measurement settings. This refinement requires an initial estimate with constant overlap with the target state. We then show how to obtain such an estimate using only $O(d)$ copies and few shallow measurement settings, so that initialization preserves the optimal dependence on $d$ in the total sample complexity. Finally, we explain how both procedures can be implemented efficiently from outcome histograms.

\subsection{Shadow Bootstrapping}
\label{sec:overview-bootstrapping}

The shadow channel $\mathcal M$ maps a state $\rho$ to the mean of the projectors $U^\dagger|b\rangle\langle b|U$ obtained from randomized measurements~\cite{huang2020predicting}.
Applying its inverse to each projector gives an unbiased shadow estimator of $\rho$ [Eq.~\eqref{eq:prelim-shadow-estimator}]. Using these estimators, shadow bootstrapping refines an initial estimate with constant overlap with the target state through successive corrections. 
We partition the measurement data into independent batches, indexed by $t=1,\ldots,J$.
At stage $t$, we use the estimate $|\sigma_{t-1}\rangle$ from the previous stage as a reference to construct $|\sigma_t\rangle$.
The purpose of this reference is to reduce fluctuations
due to the random choice of measurement bases, which cannot be removed merely by taking more shots per basis.

The key idea can be understood by analogy with
classical variance reduction.
Suppose we want to estimate $\mathbb E[X]$ and have
a prediction $Y$ whose mean is known.
Define $Z:=X-Y+\mathbb E[Y]$.
Then $\mathbb E[Z]=\mathbb E[X]$ and
\[
    \operatorname{Var}(Z)
    =\operatorname{Var}(X-Y)
    \le\mathbb E[(X-Y)^2].
\]
Thus, when $Y$ closely predicts $X$ on the same random input, $Z$ estimates the same expectation $\mathbb{E}[X]$ with small variance.

In our setting, $|\sigma_{t-1}\rangle$ predicts the outcome probabilities $|\langle z|U|\sigma_{t-1}\rangle|^2$ in each measurement basis $U$.
We subtract these predictions from the outcome frequencies observed in the experiment and use the differences to estimate the correction $\psi-\sigma_{t-1}$.
Adding back $\sigma_{t-1}$ gives an unbiased estimator of $\psi$.
Because the observed and predicted contributions use the same basis, their difference becomes less sensitive to the random choice of measurement basis as $\sigma_{t-1}$ approaches $\psi$.

Let $N_t$ denote the number of independently sampled measurement bases at stage $t$, and $S_t$ the number of shots per basis. Our analysis gives
\[
    \mathbb E\!\left[1-|\langle\psi|\sigma_t\rangle|^2\right]
    \le \frac{C}{N_t}
    \left(\mathbb E\!\left[1-|\langle\psi|\sigma_{t-1}\rangle|^2\right]+\frac{d}{S_t}\right)
\]
for a universal constant $C$. Without iterative refinement, we would take $S=\Theta(d)$ shots per basis and $N=\Theta(1/\epsilon)$ measurement bases to achieve the target accuracy $\epsilon$.
By updating the reference after each stage, we can instead choose $N_t=\Theta(1)$ and $S_t=\Theta(d\,2^t)$, with constants chosen so that the upper bound on expected infidelity is halved at each stage. After $O(\log(1/\epsilon))$ stages, the estimate therefore achieves infidelity at most $\epsilon$ with constant failure probability. 
All measurement bases and shot counts are fixed in advance; only the reference used in classical post-processing changes between stages.

Beyond pure states, we extend shadow bootstrapping to low-rank tomography with nonadaptive single-copy measurements, achieving nearly optimal bounds on both the number of measurement settings and the sample complexity (Appendix~\ref{app:sec:low-rank}).
We expect the same variance-reduction strategy to apply more broadly to channel learning and observable estimation.

\subsection{Initialization}
\label{sec:overview-initialization}

The refinement procedure requires an initial pure-state estimate $\ket{\sigma_0}$ with constant overlap with the unknown state $\ket{\psi}$. Initialization therefore need not achieve optimal dependence on accuracy. It must still use few shallow measurement settings and admit efficient processing of outcome histograms.

We use the recursive reconstruction framework of~\cite{grewal2026nearly}. At level $\ell$, each prefix $x\in\{0,1\}^{\ell}$ labels the normalized state $|\psi_x\rangle$ of the remaining $n-\ell$ qubits after measuring the first $\ell$ qubits in the computational basis and obtaining outcome $x$. Reconstruction proceeds from level $\ell=n-1$ to the root at $\ell=0$. Given normalized child estimates $|\widehat\psi_{x0}\rangle$ and $|\widehat\psi_{x1}\rangle$, we seek coefficients $\widehat\alpha_0,\widehat\alpha_1$ satisfying $|\widehat\alpha_0|^2+|\widehat\alpha_1|^2=1$ such that
\[
    |\widehat\psi_x\rangle
    =
    \widehat\alpha_0
    |0\rangle\otimes|\widehat\psi_{x0}\rangle
    +
    \widehat\alpha_1
    |1\rangle\otimes|\widehat\psi_{x1}\rangle
\]
approximates $|\psi_x\rangle$. \cite{grewal2026nearly} recover these coefficients through nested noisy searches using a Frobenius-distance oracle. Their construction achieves nearly optimal sample complexity and running time with nonadaptive Pauli measurements, using up to $\widetilde O(d/\epsilon)$ sampled measurement bases. In contrast, we recover the coefficients from linear shadow averages, allowing a small collection of shallow Clifford bases to be reused across all nodes $x\in\{0,1\}^{\ell}$ at each level $\ell$.

To describe this recovery, write the projection coefficients onto the two estimated child vectors as
\[
    \widetilde\alpha_b
    :=
    \bigl(
        \langle b|\otimes
        \langle\widehat\psi_{xb}|
    \bigr)|\psi_x\rangle,
    \qquad b\in\{0,1\}.
\]
Whenever this projection is nonzero, the normalized pair $\frac{(\widetilde\alpha_0,\widetilde\alpha_1)} {\sqrt{|\widetilde\alpha_0|^2+|\widetilde\alpha_1|^2}}$ yields the state with minimum infidelity to $|\psi_x\rangle$ within the span of the two estimated child vectors. For fixed child estimates, this pair is therefore the optimal target for $(\widehat\alpha_0,\widehat\alpha_1)$. These estimated coefficients are obtained, up to a common phase, from estimates of the following coefficient products:
\[
    \begin{pmatrix}
        |\widetilde\alpha_0|^2
        &
        \widetilde\alpha_0\widetilde\alpha_1^*
        \\
        \widetilde\alpha_0^*\widetilde\alpha_1
        &
        |\widetilde\alpha_1|^2
    \end{pmatrix}.
\]
Each entry can be expressed as an expectation value $\operatorname{Tr}(O\psi_x)$, where $\psi_x=\ket{\psi_x}\!\bra{\psi_x}$ and $O$ is determined by the estimated child states. Writing $p_x$ for the probability of observing prefix $x$, we estimate $p_x$ times each entry directly from the joint measurement outcomes. A normalized top eigenvector of the resulting $2\times2$ Hermitian matrix gives the coefficient estimates $\widehat\alpha_0,\widehat\alpha_1$.

At level $\ell$, let $N_\ell$ denote the number of independently sampled measurement bases and $S_\ell$ the number of shots per basis. Increasing $S_\ell$ reduces shot noise, but leaves uncertainty from the random choice of bases. Our analysis bounds the contribution of this uncertainty to the expected reconstruction error at each node $x$ by $O(p_x/\sqrt{N_\ell})$. Since $\sum_{x\in\{0,1\}^{\ell}}p_x=1$, summing over all nodes at level $\ell$ gives an error contribution of $O(1/\sqrt{N_\ell})$, independently of the number of nodes.

For a fixed constant target infidelity $\epsilon$ in initialization, we allocate an $O(\epsilon/n)$ error budget to the contribution from random basis selection at each level $\ell=0,\ldots,n-1$. The bound above therefore leads to the choice $N_\ell=\Theta(n^2/\epsilon^2)=\Theta(n^2)$. We choose $S_\ell$ sufficiently large at each level so that the total shot-noise contribution is $O(\epsilon)$, while keeping the total number of copies $\sum_{\ell=0}^{n-1}N_\ell S_\ell=O(d)$. Summing both contributions over the $n$ levels gives an accumulated expected infidelity of $O(\epsilon)$. The total number of sampled measurement bases is $\sum_{\ell=0}^{n-1}N_\ell=O(n^3)$, and the resulting outcome histograms can be processed in $\widetilde O(d)$ classical time.

\subsection{Post-processing from Histograms}
\label{sec:overview-postprocessing}

Given the outcome histograms, the computational task is to apply shadow estimators to state vectors without forming dense matrices. Our implementation uses two properties of the measurement ensemble. First, the shallow circuit for each sampled basis $U$ allows both $U|v\rangle$ and $U^\dagger|v\rangle$ to be computed in $\widetilde O(d)$ time. Second, the Clifford structure allows the contribution of histogram weights $h$ to be written as
\[
    \mathcal M^{-1}\!\left(
        U^\dagger\operatorname{diag}(h)U
    \right)
    =
    U^\dagger\operatorname{diag}(\lambda)U.
\]
For our ensemble, the vector $\lambda$ can be computed in $\widetilde O(d)$ time using fast Walsh--Hadamard transforms. Thus, each histogram can be processed using only $\widetilde O(d)$ time and working memory, without constructing a dense $d\times d$ estimator.

\section{Preliminaries}
\label{sec:preliminaries}

\subsection{Notation and Quantum-State Distances}

Throughout, we write $d=2^n$ and identify computational-basis outcomes with bit strings $b\in\{0,1\}^n$. The unknown pure state is denoted by $|\psi\rangle$, and $\psi=|\psi\rangle\langle\psi|$ denotes its density matrix. Similarly, we write $\sigma=|\sigma\rangle\langle\sigma|$ for a pure-state estimate. We write $\mathcal H^d:=\mathbb C^d$ for the $d$-dimensional complex Hilbert space and $\mathcal L(\mathcal H^d)$ for the space of linear operators on $\mathcal H^d$. The Hadamard ($H$), phase ($S$), and controlled-NOT (CX) gates are defined by
\[
    H:=\frac{1}{\sqrt{2}}
    \begin{pmatrix}1&1\\1&-1\end{pmatrix},
    \qquad
    S:=\begin{pmatrix}1&0\\0&i\end{pmatrix},
    \qquad
    \mathrm{CX}:=
    \begin{pmatrix}
        1&0&0&0\\
        0&1&0&0\\
        0&0&0&1\\
        0&0&1&0
    \end{pmatrix}.
\]

For vectors, $\|\cdot\|_2$ denotes the Euclidean norm. For matrices, $\|\cdot\|_{\mathrm{op}}$, $\|\cdot\|_F$, and $\|\cdot\|_{\mathrm{tr}}$ denote the operator, Frobenius, and trace norms, respectively.
For Hermitian matrices $A$ and $B$, we write $A\preceq B$ when $B-A$ is positive semidefinite.

We measure reconstruction error by the infidelity
\[
    1-\operatorname{Tr}(\psi\sigma)
    =
    1-|\langle\psi|\sigma\rangle|^2.
\]
For pure states, this is related to the Frobenius and trace distances by
\begin{equation}
\label{eq:prelim-pure-distances}
    1-\operatorname{Tr}(\psi\sigma)
    =
    \frac12\|\psi-\sigma\|_F^2
    =
    \frac14\|\psi-\sigma\|_{\mathrm{tr}}^2.
\end{equation}
In particular, the trace distance $D(\psi,\sigma)=\frac12\|\psi-\sigma\|_{\mathrm{tr}}$ satisfies the triangle inequality.

\subsection{Two-layer Clifford circuit}
\begin{figure}[H]
    \centering
    \includegraphics[width=\linewidth]{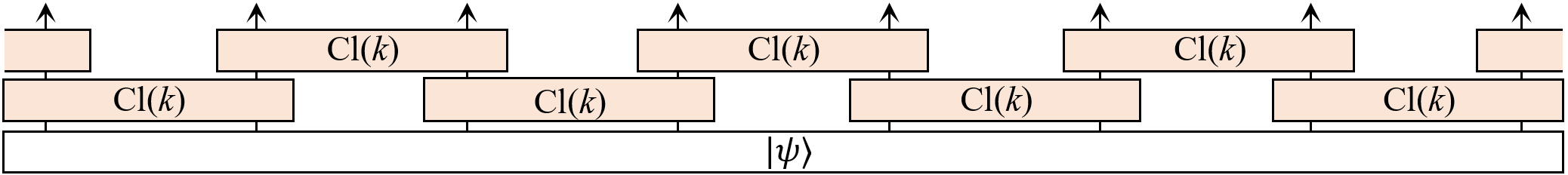}
    \caption{Two-layer Clifford circuit.}
    \label{fig:two-layer-clifford}
\end{figure}

We use the following two-layer Clifford ensemble
for our measurements.
\begin{definition}[Two-layer Clifford ensemble~\cite{schuster2025random, laracuente2026approximate}]
\label{def:two-layer-clifford} 
    Let $k$ be an even divisor of $n$. The ensemble $\mathcal E_{2\mathrm L}$ consists of circuits $U=U_2U_1$ with two staggered layers.
    Each layer is a tensor product of Clifford unitaries acting on non-overlapping contiguous $k$-qubit blocks, with periodic boundary conditions. The second-layer partition is shifted by $k/2$ qubits relative to the first.
    All block unitaries are sampled independently and uniformly from $\mathrm{Cl}(k)$. The staggered block structure is illustrated in Figure~\ref{fig:two-layer-clifford}.
\end{definition}

\subsection{Classical Shadows}

\begin{definition}[Shadow channel~\cite{huang2020predicting}]
\label{def:shadow_channel}
    Let $\mathcal E$ be an ensemble of $n$-qubit unitaries. A randomized measurement samples $U\sim\mathcal E$, applies $U$ to a copy of $\rho$, and measures in the computational basis.
    The resulting bit string $b\in\{0,1\}^n$ occurs with probability $\Pr(b\,|\,U)=\operatorname{Tr}(\rho U^{\dagger}\ket{b}\!\bra{b}U)$, and the pair $(U,b)$ forms a measurement record. The shadow channel associated with $\mathcal E$ is
    \begin{equation}
    \label{eq:prelim-shadow-channel}
        \mathcal M_{\mathcal E}(\rho)
        :=
        \mathbb E_{U\sim\mathcal E}
        \sum_b
        U^\dagger|b\rangle\langle b|U \operatorname{Tr}(\rho U^{\dagger}\ket{b}\!\bra{b}U).
    \end{equation}
\end{definition}

If $\mathcal M_{\mathcal E}$ is invertible, each record $(U,b)$ gives the shadow estimator
\begin{equation}
\label{eq:prelim-shadow-estimator}
    \hat\rho
    =
    \mathcal M_{\mathcal E}^{-1}
    \left(U^\dagger|b\rangle\langle b|U\right).
\end{equation}
By construction, $\mathbb E_{U,b}[\hat\rho]=\rho$. Consequently, for any Hermitian observable $O$, $\operatorname{Tr}(O\hat\rho)$ is an unbiased estimator of $\operatorname{Tr}(O\rho)$. We omit the subscript $\mathcal E$ when the ensemble is clear from context.

We use $P\in\{I,X,Y,Z\}^{\otimes n}$ to denote a Pauli operator. A Clifford unitary maps Pauli operators to Pauli operators up to a sign under conjugation. For a Clifford measurement ensemble, the shadow channel is diagonal in the Pauli basis~\cite{cho2025shallow}:
\begin{equation}
\label{eq:prelim-pauli-channel}
    \mathcal M_{\mathcal E}(P)=m_P P,
    \qquad
    m_P
    =
    \Pr_{U\sim\mathcal E}
    \left[
        UPU^\dagger\in\pm\{I,Z\}^{\otimes n}
    \right].
\end{equation}
We call $m_P$ the visibility of $P$. The shadow channel $\mathcal{M}_{\mathcal{E}}$ is invertible when $m_P>0$ for every Pauli operator, in which case
\begin{equation}\label{eq:inver_shadow_P}
    \mathcal M_{\mathcal E}^{-1}(P)=m_P^{-1}P.
\end{equation}

\begin{lemma}[{\cite[Appendix~G]{cho2025sample}}]
\label{lem:visibility-transfer}
    Consider the two-layer Clifford ensemble with block size $k$ in Definition~\ref{def:two-layer-clifford}, and define
    \[
        K_0:=
        \frac1{2^k+1}
        \begin{pmatrix}
            2^k+1 & 1\\
            0 & 0
        \end{pmatrix},
        \qquad
        K_1:=
        \frac1{(2^k+1)^2}
        \begin{pmatrix}
            1 & 1\\
            2^{k+1} & 2^k
        \end{pmatrix}.
    \]
    For a Pauli operator $P$, let $w_i(P)=0$ if $P$ acts as the identity on the $i$th block of the first layer, and let $w_i(P)=1$ otherwise. Then
    \[
        m_P
        =
        \operatorname{Tr}\!\left(
            K_{w_1(P)}K_{w_2(P)}
            \cdots K_{w_{n/k}(P)}
        \right).
    \]
    In particular, $m_P$ can be evaluated in $O(n)$ time.
\end{lemma}

\begin{lemma}[{Implicit in~\cite[Lemma~1]{huang2020predicting}}]
\label{lem:shadow_dual}
    For any $A,B \in \mathbb{C}^{d\times d}$,
    \begin{equation}
    \label{eq:prelim-inverse-self-adjoint}
        \operatorname{Tr}
        \left(A\,\mathcal M_{\mathcal E}^{-1}(B)\right)
        =
        \operatorname{Tr}
        \left(\mathcal M_{\mathcal E}^{-1}(A)\,B\right).
    \end{equation}
\end{lemma}

\begin{restatable}{lemma}{shadowsuminter}
\label{lem:shdow_sum_inter}
    Let $\mathcal M$ be the invertible shadow channel of a Clifford ensemble $\mathcal E$. For every $U\in\mathcal E$ and $A\in \mathbb{C}^{d\times d}$,
    \[
    \begin{aligned}
        &\sum_b
        \operatorname{Tr}
        \left(AU^\dagger|b\rangle\langle b|U\right)
        \mathcal M^{-1}\!(U^\dagger|b\rangle\langle b|U)
        =
        \sum_b
        \operatorname{Tr}
        \left(
            A\mathcal M^{-1}\!(U^\dagger|b\rangle\langle b|U)
        \right)
        U^\dagger|b\rangle\langle b|U.
    \end{aligned}
    \]
\end{restatable}

\begin{lemma}[{\cite[Appendix~A.2]{cho2025shallow}}]
\label{lem:shadow_cauchy}
Let $b,c$ be independent measurement outcomes of a fixed state conditional on the same basis $U$. For every Hermitian observable $O$,
\begin{align*}
    \mathbb E_{U,b,c}\!\left[
        \operatorname{Tr}
        \left(O\mathcal{M}^{-1}(U^{\dagger} \ket{b}\!\bra{b}U)\right)
        \operatorname{Tr}\left(O\mathcal{M}^{-1}(U^{\dagger} \ket{c}\!\bra{c}U)\right)
    \right]
    \le
    \mathbb E_{U,b}\!\left[
        \operatorname{Tr}(O\mathcal{M}^{-1}(U^{\dagger} \ket{b}\!\bra{b}U))^2
    \right].
\end{align*}
\end{lemma}

\begin{lemma}[{\cite[Appendix~D]{cho2025sample}}]\label{lem:single_shot_variance}
    For the two-layer Clifford ensemble with block size $k\ge2$ satisfying $k2^{k/2}=\Omega(n)$,
    \begin{equation}
        \|\mathbb E \hat \rho^2\|_{\operatorname{op}}=\|\mathbb E \mathcal{M}^{-1}(U^{\dagger} \ket{b}\!\bra{b}U)^2\|_{\operatorname{op}}= O(d).
    \end{equation}
\end{lemma}

\begin{lemma}[{\cite[Theorem~1]{cho2026quantum}}]
\label{lem:shadow-frobenius-variance}
    For the two-layer Clifford ensemble with block size $k\ge8$ satisfying $k2^{k/2}=\Omega(n)$, any Hermitian observable $O$ satisfies
    \begin{equation}
    \label{eq:prelim-scalar-variance}
        \operatorname{Var}[\operatorname{Tr}(O\widehat\rho)]
        \le C_F\|O\|_F^2,
        \qquad C_F=O(1).
    \end{equation}
\end{lemma}

\subsection{Histogram Representation}
\label{sec:prelim-histogram}

Let $b_1,\ldots,b_S\in\{0,1\}^n$ be independent and identically distributed measurement outcomes.
The histogram records the number of occurrences of each outcome, with empirical frequencies
\[
    \hat f_z
    =
    \frac{1}{S}\sum_{a=1}^{S}\mathbf{1}\{b_a=z\},
    \qquad
    \sum_z\hat f_z=1.
\]
We represent the experimental data as a dictionary whose keys are the observed bit strings and whose values are their corresponding counts.
Each histogram therefore contains at most $2^n$ entries. A lookup for a bit string absent from the dictionary returns zero. Our input consists of the measurement-circuit descriptions and these histograms.

\subsection{Walsh--Hadamard Transform}
\label{sec:prelim-walsh-hadamard}

For a vector $x\in\mathbb C^d$ indexed by
$i\in\{0,1\}^n$, define its Walsh--Hadamard transform $\mathrm{WH}(x) \in \mathbb C^d$ by
\[
    \mathrm{WH}(x)_j
    :=\frac{1}{2^n}
    \sum_i(-1)^{i\cdot j}x_i,
    \qquad j\in\{0,1\}^n,
\]
where $i\cdot j$ is evaluated modulo two.
Its inverse is
\[
    \mathrm{WH}^{-1}(x)_j=
    \sum_i(-1)^{i\cdot j}x_i,
    \qquad j\in\{0,1\}^n.
\]
Both transforms can be computed using $O(d\log d)$ arithmetic operations and $O(d)$ storage~\cite{fino1976unified}.

\subsection{Basic Inequalities}

We use the following two standard inequalities.

\begin{lemma}[Markov's inequality]
\label{lem:markov}
    Let $X$ be a nonnegative random variable with finite expectation. For every $a>0$,
    \[
        \Pr[X\geq a]\leq\frac{\mathbb E[X]}{a}.
    \]
\end{lemma}

\begin{lemma}[Matrix norm inequality]
\label{lem:cauchy-schwarz}
    For any random matrix $A$,
    \[
        \mathbb E\|A\|_{\mathrm{op}}
        \le \mathbb E\|A\|_F
        \le \sqrt{\mathbb E\|A\|_F^2}.
    \]
\end{lemma}

\section{Shadow Bootstrapping}
\label{sec:shadow-bootstrapping}

Shadow bootstrapping divides reconstruction into successive stages, using the current estimate as a reference to reduce the variance of the next update.
This adaptation occurs entirely in classical post-processing: all measurement bases, shot counts, and assignments to stages are fixed before any outcomes are observed.

We assume access to an initial unit vector $|\sigma_0\rangle$, satisfying
\begin{equation}\label{eq:bootstrap-warm-start}
    \operatorname{Tr}(\psi\sigma_0)
    =
    |\langle\psi|\sigma_0\rangle|^2
    \ge\frac12.
\end{equation}
Any positive constant lower bound on the overlap suffices; we choose $1/2$ for simplicity. Section~\ref{sec:initialization} constructs such an estimate using $O(d)$ copies, $\widetilde O(1)$ measurement settings, and $\widetilde O(d)$ histogram post-processing time.

\subsection{Residual estimator}
Given an unknown state $\rho$ and a reference state $\sigma$,
we define the residual as the difference between the usual shadow estimator and its counterpart constructed from the outcome probabilities predicted by $\sigma$ in the same measurement basis.
\begin{definition}[Residual estimator]
\label{def:bootstrap-residual}
    Let $\mathcal E$ be a Clifford ensemble satisfying Lemma~\ref{lem:shadow-frobenius-variance}. Fix a reference state $\sigma$ and draw $U\sim\mathcal E$. Let $\widehat f(z)$ denote the empirical frequencies obtained from $S$ independent shots $\{b_i\}_{i=1}^S$ on $\rho$ in basis $U$, as defined in Section~\ref{sec:prelim-histogram}.
    Define
    \begin{equation}
    \label{eq:bootstrap-residual}
        \Delta
        :=
        \sum_z
        \left(
            \widehat f(z)
            -
            \operatorname{Tr}\!\left(
                U\sigma U^\dagger|z\rangle\langle z|
            \right)
        \right)
        \mathcal M^{-1}
        \left(U^\dagger|z\rangle\langle z|U\right).
    \end{equation}
\end{definition}
Adding and subtracting the true outcome probabilities separates the residual into $\Delta=\Delta_{\mathrm{shot}}+\Delta_{\mathrm{floor}}$, where
\begin{align*}
    \Delta_{\mathrm{shot}}
    &:=
    \sum_z
    \left(
        \widehat f(z)
        -
        \operatorname{Tr}\!\left(
            U\rho U^\dagger|z\rangle\langle z|
        \right)
    \right)
    \mathcal M^{-1}
    \left(U^\dagger|z\rangle\langle z|U\right),\\
    \Delta_{\mathrm{floor}}
    &:=
    \sum_z
    \operatorname{Tr}\!\left(
        (\rho-\sigma)U^\dagger|z\rangle\langle z|U
    \right)
    \mathcal M^{-1}
    \left(U^\dagger|z\rangle\langle z|U\right).
\end{align*}
The term $\Delta_{\mathrm{shot}}$ captures finite-shot fluctuations around the true outcome probabilities. The term $\Delta_{\mathrm{floor}}$ fluctuates with the random choice of basis even as $S\to\infty$, setting a variance floor for fixed $\rho$ and $\sigma$. Its variation over random bases depends on the discrepancy $\rho-\sigma$ and is not suppressed by increasing $S$.

\begin{lemma}[Residual mean and variance]
\label{lem:bootstrap-residual-variance}
    For fixed states $\rho,\sigma$, the residual estimator in Definition~\ref{def:bootstrap-residual} satisfies
    \[
        \mathbb E[\Delta]=\rho-\sigma
    \]
    and
    \[
        \left\|
            \mathbb E[\Delta^2]
            -(\mathbb E[\Delta])^2
        \right\|_{\mathrm{op}}
        \le
        C\left(
            \|\rho-\sigma\|_F^2+\frac dS
        \right),\qquad C=O(1),
    \]
    where $S$ is the number of shots taken in the sampled measurement basis, and expectations are over the random basis and its $S$ measurement outcomes.
\end{lemma}
\begin{proof}
    For a fixed basis $U$, write $\texttt{b}:=\{b_i\}_{i=1}^S$ for the measurement outcomes.
    Since
    \[
        \mathbb E_{\texttt{b}}[\widehat f_z \mid U]
        =
        \operatorname{Tr}\!\left(
            \rho U^\dagger|z\rangle\langle z|U
        \right),
    \]
    $\mathbb E_{\texttt{b}}[\Delta_{\mathrm{shot}}\mid U]=0$, and hence $\mathbb E[\Delta_{\mathrm{shot}}]=0$. Moreover, by linearity and the definition of the shadow channel in Eq.~\eqref{eq:prelim-shadow-channel},
    \begin{align*}
        \mathbb E[\Delta_{\mathrm{floor}}]
        &=
        \mathbb E_U[\Delta_{\mathrm{floor}}]
        =
        \mathbb E_U\!\left[
            \sum_z
            \operatorname{Tr}\!\left(
                (\rho-\sigma)U^\dagger|z\rangle\langle z|U
            \right)
            \mathcal M^{-1}
            \!\left(U^\dagger|z\rangle\langle z|U\right)
        \right]\\
        &=
        \mathcal M^{-1}\!\left(
            \mathbb E_U\!\left[
                \sum_z
                \operatorname{Tr}\!\left(
                    (\rho-\sigma)U^\dagger|z\rangle\langle z|U
                \right)
                U^\dagger|z\rangle\langle z|U
            \right]
        \right)\\
        &=
        \mathcal M^{-1}\mathcal M(\rho-\sigma)
        =\rho-\sigma.
    \end{align*}
    Combining these identities gives
    \[
        \mathbb E[\Delta]
        =
        \mathbb E[\Delta_{\mathrm{shot}}]
        +
        \mathbb E[\Delta_{\mathrm{floor}}]
        =
        \rho-\sigma.
    \]
    Since $\Delta_{\mathrm{floor}}$ depends only on $U$,
    \begin{align*}
        \mathbb E[\Delta^2]
        &= \mathbb E[\Delta_{\mathrm{shot}}^2]
        + \mathbb E[\Delta_{\mathrm{floor}}^2]
        + \mathbb E[\Delta_{\mathrm{shot}}\Delta_{\mathrm{floor}}]
        + \mathbb E[\Delta_{\mathrm{floor}}\Delta_{\mathrm{shot}}]\\
        &= \mathbb E[\Delta_{\mathrm{shot}}^2]
        + \mathbb E[\Delta_{\mathrm{floor}}^2]
        + \mathbb E_U\!\left[
            \bigl(\mathbb E_{\texttt{b}}[\Delta_{\mathrm{shot}}\mid U]\bigr)
            \Delta_{\mathrm{floor}}
        \right]
        + \mathbb E_U\!\left[
            \Delta_{\mathrm{floor}}
            \bigl(\mathbb E_{\texttt{b}}[\Delta_{\mathrm{shot}}\mid U]\bigr)
        \right]\\
        &= \mathbb E[\Delta_{\mathrm{shot}}^2]
        + \mathbb E[\Delta_{\mathrm{floor}}^2].
    \end{align*}
    Write $\widehat\rho_{U,z}:=\mathcal M^{-1}\!\left(U^\dagger|z\rangle\langle z|U\right)$. Since the measurement outcomes from different shots are independent conditional on $U$ and $\mathbb E[\Delta_{\mathrm{shot}}\mid U]=0$,
    \begin{align*}
        \mathbb E[\Delta_{\mathrm{shot}}^2]
        &= \operatorname{Var}[\Delta_{\mathrm{shot}}]
        = \frac1S\operatorname{Var}_{U,z}\!\left[
            \widehat\rho_{U,z}
            -\mathbb E_b[\widehat\rho_{U,z}\mid U]
        \right]\\
        &= \frac1S\left(
            \mathbb E_{U,z}[\widehat\rho_{U,z}^2]
            -
            \mathbb E_U\!\left[
                \bigl(\mathbb E_z[\widehat\rho_{U,z}\mid U]\bigr)^2
            \right]
        \right)
        \preceq 
        \frac1S\mathbb E_{U,z}[\widehat\rho_{U,z}^2].
        \end{align*}
        By Lemma~\ref{lem:single_shot_variance},
        \[
            \left\|\mathbb E[\Delta_{\mathrm{shot}}^2]\right\|_{\mathrm{op}}
            \le
            \frac1S
            \left\|\mathbb E_{U,z}[\widehat\rho_{U,z}^2]\right\|_{\mathrm{op}}
            =O\left(\frac dS\right).
        \]
        By Lemma~\ref{lem:shadow_dual} and
        Lemma~\ref{lem:shdow_sum_inter},
        \begin{align*}
            \Delta_{\mathrm{floor}}
            &=
            \sum_z
            \operatorname{Tr}\!\left(
                (\rho-\sigma)U^\dagger|z\rangle\langle z|U
            \right)
            \mathcal M^{-1}\!\left(
                U^\dagger|z\rangle\langle z|U
            \right)\\
            &=
            \sum_z
            \operatorname{Tr}\!\left(
                (\rho-\sigma)\widehat\rho_{U,z}
            \right)
            U^\dagger|z\rangle\langle z|U.
        \end{align*}
        Orthogonality of the basis projectors gives
        \[
        \Delta_{\mathrm{floor}}^2
        =
        \sum_z
        \operatorname{Tr}\!\left(
            (\rho-\sigma)\widehat\rho_{U,z}
        \right)^2
        U^\dagger|z\rangle\langle z|U.
        \]
        Since $\mathbb E[\Delta_{\mathrm{floor}}]=\rho-\sigma$,
    \begin{align*}
        &\left\|
            \mathbb E[\Delta_{\mathrm{floor}}^2]
            -(\mathbb E[\Delta_{\mathrm{floor}}])^2
        \right\|_{\mathrm{op}}\\
        &=
        \mathbb E_U\!\left[
            \sum_z
            \operatorname{Tr}\!\left(
                (\rho-\sigma)\widehat\rho_{U,z}
            \right)^2 
            \operatorname{Tr}\left(\ket{v}\!\bra{v} U^{\dagger}\ket{z}\!\bra{z}U\right)
        \right]
        -\langle v|(\rho-\sigma)^2|v\rangle\\
        &=
        \mathbb E_{U,z}\!\left[
            \operatorname{Tr}\!\left(
                (\rho-\sigma)\widehat\rho_{U,z}
            \right)^2
        \right]
        -\langle v|(\rho-\sigma)^2|v\rangle\\
        &=
        \mathbb E_{U,z}\!\left[
            \operatorname{Tr}\!\left(
                (\rho-\sigma)\widehat\rho_{U,z}
            \right)^2
        \right]
        -\langle v|\rho-\sigma|v\rangle^2
        +\langle v|\rho-\sigma|v\rangle^2
        -\langle v|(\rho-\sigma)^2|v\rangle\\
        &\le
        \mathbb E_{U,z}\!\left[
            \operatorname{Tr}\!\left(
                (\rho-\sigma)\widehat\rho_{U,z}
            \right)^2
        \right]
        -\langle v|\rho-\sigma|v\rangle^2\\
        &=
        \operatorname{Var}_{U,z}\!\left[
            \operatorname{Tr}\!\left(
                (\rho-\sigma)\widehat\rho_{U,z}
            \right)
        \right]=O(\|\rho-\sigma\|_F^2),
        \tag*{(by Lemma~\ref{lem:shadow-frobenius-variance})}
    \end{align*}
    where $|v\rangle$ is a normalized eigenvector of
    $\mathbb E[\Delta_{\mathrm{floor}}^2]
    -(\mathbb E[\Delta_{\mathrm{floor}}])^2$
    with eigenvalue equal to its operator norm. Here, $z$ follows the outcome distribution obtained by applying $U$ to the auxiliary state $|v\rangle\langle v|$ and measuring in the computational basis with probability $p_z=\operatorname{Tr}(U\ket{v}\!\bra{v} U^{\dagger}\ket{z}\!\bra{z})$.
    
    Finally, using $\mathbb E[\Delta]=\rho-\sigma$, we obtain
    \begin{align*}
        \left\|
            \mathbb E[\Delta^2]-(\mathbb E[\Delta])^2
        \right\|_{\mathrm{op}}
        &=
        \left\|
            \mathbb E[\Delta_{\mathrm{shot}}^2]
            +\mathbb E[\Delta_{\mathrm{floor}}^2]
            -(\rho-\sigma)^2
        \right\|_{\mathrm{op}}\\
        &\le
        \left\|\mathbb E[\Delta_{\mathrm{shot}}^2]\right\|_{\mathrm{op}}
        +
        \left\|
            \mathbb E[\Delta_{\mathrm{floor}}^2]
            -(\rho-\sigma)^2
        \right\|_{\mathrm{op}}\\
        &\le
        C\left(\|\rho-\sigma\|_F^2+\frac dS\right),
        \qquad C=O(1).
    \end{align*}
\end{proof}

\subsection{Residual updates}

Starting from an initial estimate $|\sigma_0\rangle$, we refine the reconstruction in successive stages. At stage $t=1,\ldots,J$, we use the previous estimate $\sigma_{t-1}=|\sigma_{t-1}\rangle\langle\sigma_{t-1}|$ as a reference to construct an updated estimate $|\sigma_t\rangle$. For this update, we use $N_t$ independently sampled bases $U_{t,1},\ldots,U_{t,N_t}$, each measured on $S_t$ fresh copies, and denote their empirical frequencies by $\widehat f_{t,i}(z)$. Different stages use independent bases and disjoint copies. 

For the $i$th measurement basis $U_{t,i}$ at stage $t$, let $\Delta_{t,i}$ be the residual in Definition~\ref{def:bootstrap-residual}, with $\rho\leftarrow\psi$, $\sigma\leftarrow\sigma_{t-1}$, $U\leftarrow U_{t,i}$, and $S\leftarrow S_t$. Since $\psi$ and $\sigma_{t-1}$ are pure, $\|\psi-\sigma_{t-1}\|_F^2 =2[1-\operatorname{Tr}(\psi\sigma_{t-1})]$. Lemma~\ref{lem:bootstrap-residual-variance} therefore gives
\begin{equation}
    \mathbb E[\Delta_{t,i}\mid\sigma_{t-1}]=\psi-\sigma_{t-1},
    \qquad
    \left\|
        \mathbb E[\Delta_{t,i}^2\mid\sigma_{t-1}]
        -\bigl(\mathbb E[\Delta_{t,i}\mid\sigma_{t-1}]\bigr)^2
    \right\|_{\mathrm{op}}
    \le C\!\left(
        2[1-\operatorname{Tr}(\psi\sigma_{t-1})]+\frac d{S_t}
    \right).
\end{equation}
Thus adding the average residual to $\sigma_{t-1}$ gives an unbiased estimator of $\psi$. We apply this corrected estimator to the fixed initial vector and define 
\begin{equation}
\label{eq:bootstrap-update}
    |\widetilde\sigma_{t}\rangle
    :=
    \left(
        \sigma_{t-1}+\frac1{N_t}
        \sum_{i=1}^{N_t}\Delta_{t,i}
    \right)|\sigma_0\rangle.
\end{equation}
The next estimate is
\begin{equation}
\label{main:eq:update_rule}
    |\sigma_{t}\rangle
    :=
    \begin{cases}
        \displaystyle
        \frac{|\widetilde\sigma_{t}\rangle}
             {\||\widetilde\sigma_{t}\rangle\|_2},
        & \||\widetilde\sigma_{t}\rangle\|_2>0,\\[6pt]
        |\sigma_{t-1}\rangle,
        & \||\widetilde\sigma_{t}\rangle\|_2=0.
    \end{cases}
\end{equation}
The reference state changes between stages, while the vector $|\sigma_0\rangle$ remains fixed. To analyze the resulting update, define the residual estimation error by
\begin{equation}\label{eq:bootstrap-error}
    |e_t\rangle
    :=
    \left(
        \frac1{N_t}\sum_{i=1}^{N_t}\Delta_{t,i}
        -(\psi-\sigma_{t-1})
    \right)|\sigma_0\rangle.
\end{equation}
The update therefore satisfies the following identity
\begin{equation}\label{eq:bootstrap-signal}
    |\widetilde\sigma_{t}\rangle
    =
    \psi|\sigma_0\rangle+|e_t\rangle
    =
    \langle\psi|\sigma_0\rangle|\psi\rangle+|e_t\rangle.
\end{equation}
As $\|\!\ket{e_t}\!\|_2\to0$, $\ket{\widetilde\sigma_{t}}$ approaches a nonzero multiple of $\ket{\psi}$. After normalization, the estimate converges to the desired state. The following lemma bounds the expected squared norm of $\ket{e_t}$ at each stage.

\begin{lemma}[Residual estimation error]
\label{lem:bootstrap-residual-error}
    For a fixed current estimate $\sigma_{t-1}$, the expectation over the measurement bases and outcomes at stage $t$ satisfies
    \[
    \begin{aligned}
        \mathbb E[\|\!\ket{e_t}\!\|_2^2 \mid \sigma_{t-1}]
        \le
        \frac{2C}{N_t}
        \left(
            1-\operatorname{Tr}(\psi\sigma_{t-1})
            +\frac d{2S_t}
        \right).
    \end{aligned}
    \]
\end{lemma}
\begin{proof}
Conditioned on the current estimate $\sigma_{t-1}$, the residuals $\Delta_{t,i}$ are independent and satisfy $\mathbb E[\Delta_{t,i}\mid\sigma_{t-1}]=\psi-\sigma_{t-1}$ by Lemma~\ref{lem:bootstrap-residual-variance}. The centered cross terms therefore have zero conditional expectation, giving
\begin{align*}
    \mathbb E[\|\!\ket{e_t}\!\|_2^2\mid\sigma_{t-1}]
    &=
    \frac1{N_t}
    \langle\sigma_0|
    \left(
        \mathbb E[\Delta_{t,1}^2\mid\sigma_{t-1}]
        -(\psi-\sigma_{t-1})^2
    \right)
    |\sigma_0\rangle\\
    &\le
    \frac1{N_t}
    \left\|
        \mathbb E[\Delta_{t,1}^2\mid\sigma_{t-1}]
        -(\psi-\sigma_{t-1})^2
    \right\|_{\mathrm{op}}\\
    &\le
    \frac{C}{N_t}
    \left(
        \|\psi-\sigma_{t-1}\|_F^2+\frac d{S_t}
    \right)
    \tag*{(by Lemma~\ref{lem:bootstrap-residual-variance})}\\
    &=
    \frac{2C}{N_t}
    \left(
        1-\operatorname{Tr}(\psi\sigma_{t-1})
        +\frac d{2S_t}
    \right).
\end{align*}
\end{proof}

\begin{lemma}[From vector error to infidelity]
\label{lem:bootstrap-error-conversion}
    Given $\operatorname{Tr}(\psi\sigma_0)\ge1/2$,
    \begin{equation}
        1-\operatorname{Tr}(\psi\sigma_{t})
        \le
        \frac{\|\!\ket{e_t}\!\|_2^2}{\operatorname{Tr}(\psi\sigma_0)}
        \le
        2\|\!\ket{e_t}\!\|_2^2. \label{eq:fid_error}
    \end{equation}
\end{lemma}

\begin{proof}
    The update rule in Eq.~\eqref{main:eq:update_rule} ensures
    \[
        (I-\sigma_{t})|\widetilde\sigma_{t}\rangle=0,
    \]
    including when $\||\widetilde\sigma_{t}\rangle\|_2=0$.
    Using Eq.~\eqref{eq:bootstrap-signal}, we obtain
    \begin{align*}
        &(I-\sigma_{t})\ket{\tilde \sigma_{t}} = 0
        \\
        \quad&\implies 
        (I-\sigma_{t})\psi|\sigma_0\rangle = -(I-\sigma_{t})|e_t\rangle
        \\
        \quad&\implies
        (I-\sigma_{t})\langle \psi|\sigma_0\rangle |\psi\rangle = -(I-\sigma_{t})|e_t\rangle
        \\
        \quad&\implies
        |\langle \psi|\sigma_0\rangle|^2 \|(I-\sigma_{t})|\psi\rangle\|_2^2 = \|(I-\sigma_{t})|e_t\rangle\|_2^2
        \\
        \quad&\implies
        \operatorname{Tr}(\psi \sigma_0) (1-\operatorname{Tr}(\psi\sigma_{t} )) = \bra{e_t}(I-\sigma_{t})\ket{e_t} \le \|\!\ket{e_t}\!\|_2^2
    \end{align*}
    Dividing by $\operatorname{Tr}(\psi\sigma_0)\ge1/2$
    proves the claim.
\end{proof}

If the initial vector $\ket{\sigma_0}$ is chosen uniformly at random, its expected squared overlap with the target is only $1/d=2^{-n}$. Since the bound in Eq.~\eqref{eq:fid_error} divides by $\operatorname{Tr}(\psi\sigma_0)$, a small overlap can greatly weaken the infidelity guarantee.

\begin{corollary}
\label{cor:bootstrap-convergence}
    Fix an initial estimate $\sigma_0$ with
    $\operatorname{Tr}(\psi\sigma_0)\ge 1/2$.
    For any $\epsilon\in(0,1)$, let
    $J:=\lceil\log_2(1/(2\epsilon))\rceil$.
    There exist choices of $N_t,S_t$ for $t=1,\ldots,J$
    such that
    $\mathbb E[1-\operatorname{Tr}(\psi\sigma_J)]\le\epsilon$
    and
    \begin{equation*}
        \sum_{t=1}^{J}N_t=O(\log(1/\epsilon)),
        \qquad
        \sum_{t=1}^{J}N_tS_t=O(d/\epsilon).
    \end{equation*}
\end{corollary}

\begin{proof}
    Choose $N_t:=\lceil16(C+1)\rceil$ and $S_t:=d\,2^{t-1}$ for $t=1,\ldots,J$. By Lemmas~\ref{lem:bootstrap-residual-error} and~\ref{lem:bootstrap-error-conversion}, with all previous stages held fixed,
    \begin{align*}
        \mathbb E\left[1-\operatorname{Tr}(\psi\sigma_{t})
            \mid\sigma_{t-1}\right]
        &\le 2\,\mathbb E[\|\!\ket{e_t}\!\|_2^2\mid\sigma_{t-1}]\\
        &\le \frac{4C}{N_t}
            \left(
                1-\operatorname{Tr}(\psi\sigma_{t-1})
                +\frac{d}{2S_t}
            \right)\\
        &\le \frac14
            [1-\operatorname{Tr}(\psi\sigma_{t-1})]
            +\frac{d}{8S_t}.
    \end{align*}
    The initial error is at most $1/2$.
    If $\mathbb E[1-\operatorname{Tr}(\psi\sigma_{t-1})]
    \le2^{-t}$, then averaging over previous stages gives
    \begin{align*}
        \mathbb E[1-\operatorname{Tr}(\psi\sigma_{t})]
        &\le
        \frac14\,\mathbb E[1-\operatorname{Tr}(\psi\sigma_{t-1})]
        +\frac{d}{8S_t}\\
        &\le
        \frac{1}{4\cdot2^{t}}
        +\frac{1}{8\cdot2^{t-1}}
        =2^{-(t+1)}.
    \end{align*}
    Induction therefore gives
    $\mathbb E[1-\operatorname{Tr}(\psi\sigma_J)]
    \le2^{-(J+1)}\le\epsilon$.
    
    Finally, $N_t=O(1)$ and $J=O(\log(1/\epsilon))$, so
    \begin{align*}
        \sum_{t=1}^{J}N_t
        &=O(J)=O(\log(1/\epsilon)),\\
        \sum_{t=1}^{J}N_tS_t
        &=O\!\left(d\sum_{t=1}^{J}2^t\right)
          =O\!\left(d\,2^J\right)
          =O(d/\epsilon).
    \end{align*}
\end{proof}

\begin{corollary}[Constant success probability]
\label{cor:bootstrap-success}
    Fix an initial estimate $\sigma_0$ with $\operatorname{Tr}(\psi\sigma_0)\ge1/2$. For any $0<\epsilon<1$, running the refinement with expected-error target $\epsilon/12$ yields
    \begin{equation*}
        \Pr[1-\operatorname{Tr}(\psi\sigma_J)>\epsilon]
        \le\frac1{12}.
    \end{equation*}
    The asymptotic resource bounds of Corollary~\ref{cor:bootstrap-convergence} remain unchanged.
\end{corollary}

\begin{proof}
    By Corollary~\ref{cor:bootstrap-convergence} and Markov's inequality,
    \begin{align*}
        \Pr[1-\operatorname{Tr}(\psi\sigma_J)>\epsilon]
        &\le
        \frac{\mathbb E[1-\operatorname{Tr}(\psi\sigma_J)]}
             {\epsilon}\\
        &\le \frac{\epsilon/12}{\epsilon}
        =\frac1{12}.
    \end{align*}
    Replacing $\epsilon$ by $\epsilon/12$ increases the number of stages by only an additive constant and preserves the asymptotic resource bounds.
\end{proof}

\section{Initialization}
\label{sec:initialization}

We construct a normalized estimate $|\sigma_0\rangle$ with constant overlap with an unknown pure state $|\psi\rangle\in\mathbb C^d$, where $d=2^n$, to serve as the starting point for shadow bootstrapping. The procedure uses nonadaptive shallow Clifford measurements and recursively combines estimates of smaller conditional states. The following theorem summarizes its guarantees.
\begin{theorem}[Initialization]
\label{thm:initialization}
    Under the measurement assumptions above, there are choices of
    $N_\ell$ and $S_\ell$ such that the reconstruction below outputs
    a unit vector $|\sigma_0\rangle$ satisfying
    \[
        \Pr\!\left[
            |\langle\psi|\sigma_0\rangle|^2\ge\frac12
        \right]\ge\frac{11}{12}.
    \]
    The protocol is nonadaptive and single-copy, with
    \[
        \sum_{\ell=0}^{n-1}N_\ell S_\ell=O(d),
        \qquad
        \sum_{\ell=0}^{n-1}N_\ell=O(n^3).
    \]
    Each measurement has Clifford depth $O(\log n)$.
\end{theorem}

\subsection{Review of the tree representation of $\ket{\psi}$} 
The reconstruction proceeds along a binary tree whose nodes at level $\ell$ are indexed by prefixes $x\in\{0,1\}^{\ell}$.

\begin{definition}[Conditional states]
\label{def:init-conditional-states}
    For a prefix $x\in\{0,1\}^{\ell}$, define $p_x:=\bigl\|(\langle x|\otimes I)|\psi\rangle\bigr\|_2^2$.
    If $p_x>0$, define the normalized conditional state
    \[
        |\psi_x\rangle
        :=
        \frac{(\langle x|\otimes I)|\psi\rangle}{\sqrt{p_x}}
        \in\mathbb C^{2^{n-\ell}},
        \qquad
        \psi_x:=|\psi_x\rangle\langle\psi_x|.
    \]
    At the root, $p_{\varnothing}=1$ and $|\psi_{\varnothing}\rangle=|\psi\rangle$.
\end{definition}

\begin{lemma}[{\cite[The proof of Lemma~4]{grewal2026nearly}}]
    For every $\ell=0,\ldots,n-1$ and prefix $x\in\{0,1\}^{\ell}$, the parent and child states satisfy
    \begin{equation}
        \label{eq:init-parent-child}
        \begin{aligned}
            p_x
            &=p_{x0}+p_{x1},\\
            \sqrt{p_x}\,|\psi_x\rangle
            &=
            \sqrt{p_{x0}}\,|0\rangle\otimes|\psi_{x0}\rangle
            +
            \sqrt{p_{x1}}\,|1\rangle\otimes|\psi_{x1}\rangle.
        \end{aligned}
    \end{equation}
    In particular, $\sum_{x\in\{0,1\}^{\ell}}p_x=1$ at every level.
\end{lemma}

\subsection{Measurement setup}

\begin{lemma}[Informal, see Lemma~\ref{appx:lem:suffix-ensembles}]
\label{lem:suffix-ensembles}
    For each $\ell=0,\ldots,n-1$, there exists an $O(\log n)$-depth Clifford ensemble $\mathcal E_\ell$ on $n-\ell$ qubits with a shadow channel $\mathcal M_\ell$ satisfying Eq.~\eqref{eq:prelim-scalar-variance} for every state and Hermitian observable, with a universal constant.
\end{lemma}

\begin{figure}[H]
    \centering
    \includegraphics[width=\linewidth]{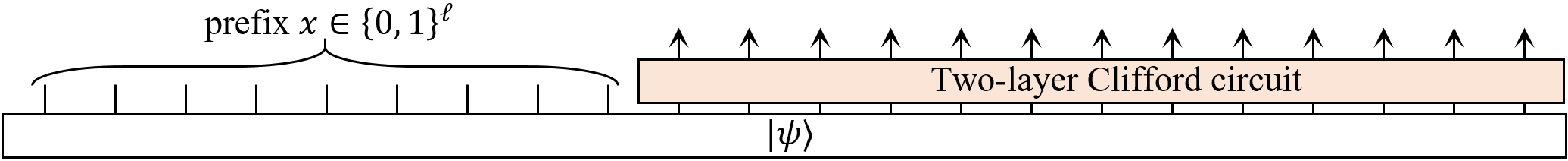}
    \caption{Suffix unitary ensemble $\mathcal E_{\ell}$.}
    \label{fig:suffix-unitary-ensemble}
\end{figure}

At each level $\ell$, independently sample
\[
    U_1,\ldots,U_{N_\ell}\sim\mathcal E_\ell.
\]
We suppress the level index in $U_i$. The bases sampled at different levels are independent. For each sampled $U_{i}$, apply $I_{2^\ell}\otimes U_{i}$ to $S_\ell$ copies and measure all qubits in the computational basis. As illustrated in Figure~\ref{fig:suffix-unitary-ensemble}, the first $\ell$ qubits are measured in the $Z$ basis, while the two-layer Clifford circuit acts only on the remaining $n-\ell$ qubits.

Level $\ell$ therefore uses $N_\ell$ sampled bases and $N_\ell S_\ell$ copies. All bases and shot counts are chosen in advance, and different levels use independent bases and disjoint copies. We write a measurement outcome as
\[
    (x,b)\in
    \{0,1\}^{\ell}\times\{0,1\}^{n-\ell},
\]
where $x$ is the $\ell$-bit prefix and $b$ is the $(n-\ell)$-bit suffix. For a fixed basis $U$, this outcome occurs with probability
\begin{equation}\label{eq:init-outcome-distribution}
    \Pr(x,b)
    =
    p_x\,\langle b|U\psi_xU^\dagger|b\rangle.
\end{equation}

\subsection{Recursive reconstruction}
The reconstruction proceeds from the leaves to the root. Suppose that normalized child estimates $|\widehat\psi_{x0}\rangle$ and $|\widehat\psi_{x1}\rangle$ have already been constructed. For $p_x>0$, the true parent state can be decomposed as
\begin{align*}
    |\psi_x\rangle
    &=
    \sqrt{\frac{p_{x0}}{p_x}}\,
        |0\rangle\otimes|\psi_{x0}\rangle
    +
    \sqrt{\frac{p_{x1}}{p_x}}\,
        |1\rangle\otimes|\psi_{x1}\rangle \label{eq:decomp_exact}
    \\
    &=
    \widetilde \alpha_0\,
        |0\rangle\otimes|\widehat\psi_{x0}\rangle
    +
    \widetilde \alpha_1\,
        |1\rangle\otimes|\widehat\psi_{x1}\rangle
    +
    \text{(an orthogonal remainder)},
\end{align*}
where $\widetilde\alpha_i := (\langle i|\otimes\langle\widehat\psi_{xi}|)|\psi_x\rangle$ for $i\in\{0,1\}$, and the remainder is orthogonal to both $|0\rangle\otimes|\widehat\psi_{x0}\rangle$ and $|1\rangle\otimes|\widehat\psi_{x1}\rangle$. Whenever $|\widetilde\alpha_0|^2+|\widetilde\alpha_1|^2>0$, the normalized projection onto their span has coefficient vector $\alpha:=(\alpha_0,\alpha_1)^{\mathsf T}\in\mathbb C^2$, where
\[
    \alpha_0
    =
    \frac{\widetilde \alpha_0}
         {\sqrt{|\widetilde \alpha_0|^2+|\widetilde \alpha_1|^2}},
    \qquad
    \alpha_1
    =
    \frac{\widetilde \alpha_1}
         {\sqrt{|\widetilde \alpha_0|^2+|\widetilde \alpha_1|^2}}.
\]
Our goal is to estimate $\alpha$ up to a common phase. Repeating this reconstruction up the tree yields the estimate at the root. 

We adopt the recursive reconstruction framework of~\cite{grewal2026nearly}, but replace their nested one-dimensional noisy searches with direct estimation of a $2\times2$ reconstruction matrix using shallow Clifford measurements. Its entries are $p_x$ times the following coefficient products:
\begin{align*}
    |\tilde\alpha_0|^2
    &=
    \operatorname{Tr}\!\left[
        \left(
            |0\rangle\langle0|
            \otimes
            |\hat\psi_{x0}\rangle\langle\hat\psi_{x0}|
        \right)\psi_x
    \right],
    \qquad
    \tilde\alpha_0\tilde\alpha_1^*
    =
    \operatorname{Tr}\!\left[
        \left(
            |1\rangle\langle0|
            \otimes
            |\hat\psi_{x1}\rangle\langle\hat\psi_{x0}|
        \right)\psi_x
    \right],
    \\
    \tilde\alpha_0^*\tilde\alpha_1
    &=
    \operatorname{Tr}\!\left[
        \left(
            |0\rangle\langle1|
            \otimes
            |\hat\psi_{x0}\rangle\langle\hat\psi_{x1}|
        \right)\psi_x
    \right],
    \qquad
    |\tilde\alpha_1|^2
    =
    \operatorname{Tr}\!\left[
        \left(
            |1\rangle\langle1|
            \otimes
            |\hat\psi_{x1}\rangle\langle\hat\psi_{x1}|
        \right)\psi_x
    \right].
\end{align*}

Fix a node $x\in\{0,1\}^{\ell}$ with normalized child estimates, and write
\[
    |q_0\rangle
    :=|0\rangle\otimes|\widehat\psi_{x0}\rangle,
    \qquad
    |q_1\rangle
    :=|1\rangle\otimes|\widehat\psi_{x1}\rangle.
\]
These vectors are orthonormal.
For $w=(w_0,w_1)\in\mathbb C^2$ with $\|w\|_2=1$, define
\[
    |q_w\rangle:=w_0|q_0\rangle+w_1|q_1\rangle.
\]

\begin{definition}[Reconstruction matrix]
\label{def:init-core}
    Define $B_x\in\mathbb C^{2\times2}$ by
    \[
        (B_x)_{ab}
        :=p_x\langle q_a|\psi_x\rangle
                 \langle\psi_x|q_b\rangle,
        \qquad a,b\in\{0,1\}.
    \]
    The factor $p_x$ allows the entries to be estimated unbiasedly from the measurement data without dividing by the unknown prefix probability.
\end{definition}

\begin{lemma}
\label{lem:init-quadratic-form}
    The matrix $B_x$ is positive semidefinite with rank at most one. For every unit vector $w\in\mathbb C^2$, $w^\dagger B_xw=p_x|\langle\psi_x|q_w\rangle|^2.$
\end{lemma}
\begin{proof}
    By Definition~\ref{def:init-core},
    \[
        B_x
        =
        p_x
        \begin{pmatrix}
            |\widetilde\alpha_0|^2
            & \widetilde\alpha_0\widetilde\alpha_1^*\\
            \widetilde\alpha_0^*\widetilde\alpha_1
            & |\widetilde\alpha_1|^2
        \end{pmatrix}
        =
        p_x
        \begin{pmatrix}
            \widetilde\alpha_0\\
            \widetilde\alpha_1
        \end{pmatrix}
        \bigl(\widetilde\alpha_0^*,\,\widetilde\alpha_1^*\bigr).
    \]
    If $|\psi_x\rangle$ is orthogonal to both $|q_0\rangle$ and $|q_1\rangle$, then $\widetilde\alpha_0=\widetilde\alpha_1=0$ and $B_x=0$. Otherwise,
    \[
        B_x
        =
        p_x\bigl(
            |\widetilde\alpha_0|^2+|\widetilde\alpha_1|^2
        \bigr)\alpha\alpha^\dagger.
    \]
    In either case, $B_x$ is positive semidefinite with rank at most one. 
    
    For every unit vector $w\in\mathbb C^2$,
    \begin{align*}
        w^\dagger B_xw
        &=
        p_x\sum_{a,b\in\{0,1\}}
        w_a^*w_b
        \langle q_a|\psi_x\rangle
        \langle\psi_x|q_b\rangle\\
        &=
        p_x\left|
            \sum_{b\in\{0,1\}}w_b\langle\psi_x|q_b\rangle
        \right|^2=p_x|\langle\psi_x|q_w\rangle|^2.
    \end{align*}
\end{proof}

\begin{corollary}[Inherited reconstruction error]
\label{lem:init-inherited-error}
    For every node $x$,
    \begin{align}
        p_x-\alpha^{\dagger} B_x \alpha
        =
        p_{x0}\bigl(
            1-|\langle\psi_{x0}|\widehat\psi_{x0}\rangle|^2
        \bigr)+
        p_{x1}\bigl(
            1-|\langle\psi_{x1}|\widehat\psi_{x1}\rangle|^2
        \bigr).
    \end{align}
\end{corollary}
\begin{proof}
    Recall that $\alpha$ is the normalized projection coefficient vector defined above. By Lemma~\ref{lem:init-quadratic-form},
    \begin{align*}
        \alpha^\dagger B_x\alpha
        &=
        p_x\bigl(
            |\widetilde\alpha_0|^2+|\widetilde\alpha_1|^2
        \bigr)\\
        &=
        p_{x0}|\langle\psi_{x0}|\widehat\psi_{x0}\rangle|^2
        +
        p_{x1}|\langle\psi_{x1}|\widehat\psi_{x1}\rangle|^2.
    \end{align*}
    Subtracting this from $p_x=p_{x0}+p_{x1}$ proves the claim.
\end{proof}

\subsection{Conditional shadow estimator}

\begin{definition}[Conditional shadow estimator]
    Fix a level $\ell$ and a node $x\in\{0,1\}^{\ell}$. For each $i=1,\ldots,N_\ell$, let $\{(X_{i,a},b_{i,a})\}_{a=1}^{S_\ell}$ denote the $S_\ell$ measurement outcomes in basis $U_i$. For a Hermitian suffix observable $O$, define
    \begin{equation}
    \label{eq:init-observable-estimator}
        \begin{aligned}
            \widehat o
            &:=
            \frac1{N_\ell}\sum_{i=1}^{N_\ell}
            \widehat o_{x,U_i},\\
            \widehat o_{x,U_i}
            &:=
            \frac1{S_\ell}\sum_{a=1}^{S_\ell}
            \mathbf1_{\{X_{i,a}=x\}}
            \operatorname{Tr}\!\left[
                O\mathcal M_\ell^{-1}
                \bigl(
                    U_i^\dagger
                    |b_{i,a}\rangle\langle b_{i,a}|
                    U_i
                \bigr)
            \right].
        \end{aligned}
    \end{equation}
    When several observables are considered, write $\widehat o(O)$ for the corresponding estimator. Note that each basis average $\hat{o}_{x,U}$ is normalized by all $S_\ell$ shots.
\end{definition}

\begin{lemma}
\label{lem:init-core-variance}
    Fix $x\in\{0,1\}^{\ell}$. For any Hermitian observable $O\in\mathcal L(\mathcal H^{2^{n-\ell}})$ on the suffix, the estimator in Eq.~\eqref{eq:init-observable-estimator} satisfies
    \[
        \mathbb E[\widehat o]
        =p_x\operatorname{Tr}(O\psi_x)
    \]
    and
    \[
        \operatorname{Var}[\widehat o]
        \le
        \frac{C\|O\|_F^2}{N_\ell}
        \left(p_x^2+\frac{p_x}{S_\ell}\right).
    \]
\end{lemma}
\begin{proof}
    Take expectations over the level-$\ell$ measurement data. Using the measurement probabilities in Eq.~\eqref{eq:init-outcome-distribution}, we obtain
    \begin{align*}
        \mathbb E[\widehat o]
        &=
        \frac1{N_\ell}\sum_{i=1}^{N_\ell}
        \mathbb E[\widehat o_{x,U_i}]\\
        &=
        p_x\,\mathbb E_U\!\left[
            \sum_b
            \operatorname{Tr}\!\left(
                \psi_xU^\dagger|b\rangle\langle b|U
            \right)
            \operatorname{Tr}\!\left(
                O\mathcal M_\ell^{-1}
                (U^\dagger|b\rangle\langle b|U)
            \right)
        \right]\\
        &=
        p_x\operatorname{Tr}\!\left(
            O\mathcal M_\ell^{-1}\mathcal M_\ell(\psi_x)
        \right)\\
        &=p_x\operatorname{Tr}(O\psi_x).
    \end{align*}
    Independence across the sampled bases and their associated measurement outcomes gives
    \begin{align*}
        \operatorname{Var}[\widehat o]
        &=
        \frac1{N_\ell}\operatorname{Var}[\widehat o_{x,U}]\\
        &=
        \frac1{N_\ell}
        \left(
        \mathbb{E}[\widehat o_{x,U}^2]-p_x^2 \operatorname{Tr}(O\psi_x)^2
        \right)\\
        &=
        \frac1{N_\ell}
        \left(\frac{p_x}{S_\ell}
        \mathbb{E}_{U,b}[\operatorname{Tr}(O\widehat\rho_{U,b})^2] 
        + 
        \frac{p_x^2(S_{\ell}-1)}{S_\ell}
        \mathbb{E}_{U,b,c}[\operatorname{Tr}(O\widehat\rho_{U,b})\operatorname{Tr}(O\widehat\rho_{U,c})]
        -p_x^2 \operatorname{Tr}(O\psi_x)^2
        \right)\\
        &=
        \frac1{N_\ell}\Bigg(
            \frac{p_x}{S_\ell}
            \operatorname{Var}_{U,b}
            [\operatorname{Tr}(O\widehat\rho_{U,b})]
            +
            \frac{p_x(1-p_x)}{S_\ell}
            \operatorname{Tr}(O\psi_x)^2\\
        &\qquad+
            \frac{p_x^2(S_\ell-1)}{S_\ell}
            \left(
                \mathbb E_{U,b,c}\!\left[
                    \operatorname{Tr}(O\widehat\rho_{U,b})
                    \operatorname{Tr}(O\widehat\rho_{U,c})
                \right]
                -\operatorname{Tr}(O\psi_x)^2
            \right)
        \Bigg),
    \end{align*}
    where $b,c$ are independent outcomes of measuring $\psi_x$ conditional on $U$.
    By Lemma~\ref{lem:shadow_cauchy},
    \[
        \mathbb E_{U,b,c}\!\left[
            \operatorname{Tr}(O\widehat\rho_{U,b})
            \operatorname{Tr}(O\widehat\rho_{U,c})
        \right]
        -\operatorname{Tr}(O\psi_x)^2
        \le
        \operatorname{Var}_{U,b}
        [\operatorname{Tr}(O\widehat\rho_{U,b})].
    \]
    Applying Lemma~\ref{lem:shadow-frobenius-variance} and $\operatorname{Tr}(O\psi_x)^2\le\|O\|_F^2$ gives
    \begin{align*}
        \operatorname{Var}[\widehat o]
        &\le
        \frac{C\|O\|_F^2}{N_\ell}
        \left(
        \frac{p_x^2(S_\ell-1)}{S_\ell}+
            \frac{p_x}{S_\ell}
        \right)\\
        &\le
        \frac{C\|O\|_F^2}{N_\ell}
        \left(p_x^2+\frac{p_x}{S_\ell}\right).
    \end{align*}
\end{proof}

\begin{restatable}[Complex observables]{corollary}{initcomplexvariance}
\label{cor:init-complex-variance}
    Fix $x\in\{0,1\}^{\ell}$. For any observable $O\in\mathcal L(\mathcal H^{2^{n-\ell}})$ on the suffix, the estimator in Eq.~\eqref{eq:init-observable-estimator} satisfies $\mathbb E[\widehat o]=p_x\operatorname{Tr}(O\psi_x)$ and
    \[
        \mathbb E\!\left[
            \left|\widehat o-p_x\operatorname{Tr}(O\psi_x)\right|^2
        \right]
        \le
        \frac{C\|O\|_F^2}{N_\ell}
        \left(p_x^2+\frac{p_x}{S_\ell}\right).
    \]
\end{restatable}

\subsection{Estimating $B_x$}
Define
\[
    \begin{aligned}
        O_0&:=|q_0\rangle\langle q_0|,
        &
        O_1&:=|q_1\rangle\langle q_0|,\\
        O_2&:=|q_0\rangle\langle q_1|,
        &
        O_3&:=|q_1\rangle\langle q_1|.
    \end{aligned}
\]
Using the estimator in Eq.~\eqref{eq:init-observable-estimator}, define
\begin{equation}
    \widehat B_x
    :=
    \begin{pmatrix}
        \widehat o(O_0) & \widehat o(O_1)\\
        \widehat o(O_2) & \widehat o(O_3)
    \end{pmatrix}.
\end{equation}
This matrix is an unbiased estimator of $B_x$. All four estimates use the same measurement data at level $\ell$. Since the shadow estimators are Hermitian and $O_2=O_1^\dagger$, the resulting matrix $\widehat B_x$ is Hermitian.

Diagonalize the $2\times2$ matrix $\widehat B_x$, and let $\widehat\alpha=(\widehat\alpha_0,\widehat\alpha_1)^{\mathsf T}$ be a normalized eigenvector corresponding to its largest eigenvalue. If this eigenvalue is repeated, choose any normalized vector in its eigenspace. We use $\widehat\alpha$ as an estimate of $\alpha$ and define the normalized parent estimate by
\begin{equation}\label{eq:est_prefix_psi}
        |\widehat\psi_x\rangle
    :=
    |q_{\widehat\alpha}\rangle
    =
    \widehat\alpha_0\,
        |0\rangle\otimes|\widehat\psi_{x0}\rangle
    +
    \widehat\alpha_1\,
        |1\rangle\otimes|\widehat\psi_{x1}\rangle.
\end{equation}
When $\widehat B_x=B_x$, Corollary~\ref{lem:init-inherited-error} shows that this reconstruction inherits only errors from its child nodes. The following lemma bounds the additional error caused by estimating $B_x$.

\begin{lemma}[Error propagation at a node]
\label{lem:init-gluing}
    For every node $x$, normalized child estimates $|\widehat\psi_{x0}\rangle, |\widehat\psi_{x1}\rangle$, and Hermitian estimate $\widehat B_x$, the reconstruction above satisfies
    \[
    \begin{aligned}
        p_x\bigl(1-|\langle\psi_x|\widehat\psi_x\rangle|^2\bigr)
        \le{}&
        \underbrace{
            p_{x0}\bigl(
                1-|\langle\psi_{x0}|\widehat\psi_{x0}\rangle|^2
            \bigr)
            +
            p_{x1}\bigl(
                1-|\langle\psi_{x1}|\widehat\psi_{x1}\rangle|^2
            \bigr)
        }_{\text{Error from the child nodes}}
        \\[4pt]
        &+
        \underbrace{
            2\|\widehat B_x-B_x\|_{\mathrm{op}}
        }_{\text{Error at the current node}}.
    \end{aligned}
    \]
\end{lemma}

\begin{proof}
Set $E_x:=\widehat B_x-B_x$. Then
\begin{align*}
    p_x\bigl(1-|\langle\psi_x|\widehat\psi_x\rangle|^2\bigr)
    &=p_x\bigl(1-|\langle\psi_x|q_{\widehat{\alpha}}\rangle|^2\bigr)
    \tag*{(by Eq.~\eqref{eq:est_prefix_psi})}\\
    &=
    p_x-\widehat\alpha^\dagger B_x\widehat\alpha
    \tag*{(by Lemma~\ref{lem:init-quadratic-form})}\\
    &=
    p_x-\alpha^\dagger B_x\alpha
    +\alpha^\dagger B_x\alpha
    -\widehat\alpha^\dagger B_x\widehat\alpha\\
    &=
    p_x-\alpha^\dagger B_x\alpha
    +\alpha^\dagger\widehat B_x\alpha
    -\alpha^\dagger E_x\alpha
    -\widehat\alpha^\dagger B_x\widehat\alpha\\
    &\le
    p_x-\alpha^\dagger B_x\alpha
    +\widehat\alpha^\dagger\widehat B_x\widehat\alpha
    -\alpha^\dagger E_x\alpha
    -\widehat\alpha^\dagger B_x\widehat\alpha
    \tag*{(by maximality of $\widehat\alpha$)}\\
    &=
    p_x-\alpha^\dagger B_x\alpha
    +\widehat\alpha^\dagger E_x\widehat\alpha
    -\alpha^\dagger E_x\alpha\\
    &\le
    p_x-\alpha^\dagger B_x\alpha
    +2\|E_x\|_{\mathrm{op}}
    \tag*{(since $\|\alpha\|_2=\|\widehat\alpha\|_2=1$)}\\
    &=
    \sum_{b\in\{0,1\}}
    p_{xb}\bigl(
        1-|\langle\psi_{xb}|\widehat\psi_{xb}\rangle|^2
    \bigr)
    +2\|\widehat B_x-B_x\|_{\mathrm{op}}.
    \tag*{(by Corollary~\ref{lem:init-inherited-error})}
\end{align*}
\end{proof}

\begin{lemma}
\label{lem:init-core-estimation}
    For every node $x\in\{0,1\}^{\ell}$ and fixed child estimates, the expectation over the independent measurement data at level $\ell$ satisfies
    \[
        \mathbb E[\widehat B_x]=B_x
    \]
    and
    \[
        \mathbb E\!\left[
            \|\widehat B_x-B_x\|_F^2
        \right]
        \le
        \frac{4C}{N_\ell}
        \left(p_x^2+\frac{p_x}{S_\ell}\right).
    \]
\end{lemma}
\begin{proof}
    With the child estimates held fixed, the observables $O_0,\ldots,O_3$ are fixed and satisfy $\|O_j\|_F=1$. By the definitions of $B_x$ and $\widehat B_x$, Corollary~\ref{cor:init-complex-variance} applied entrywise gives $\mathbb E[\widehat B_x]=B_x$. Moreover,
    \begin{align*}
        \mathbb E\!\left[\|\widehat B_x-B_x\|_F^2\right]
        &=
        \sum_{j=0}^3
        \mathbb E\!\left[
            \left|
                \widehat o(O_j)-p_x\operatorname{Tr}(O_j\psi_x)
            \right|^2
        \right]\\
        &\le
        \frac{C}{N_\ell}
        \left(p_x^2+\frac{p_x}{S_\ell}\right)
        \sum_{j=0}^3\|O_j\|_F^2
        \tag{by Corollary~\ref{cor:init-complex-variance}}
        \\
        &=
        \frac{4C}{N_\ell}
        \left(p_x^2+\frac{p_x}{S_\ell}\right).
    \end{align*}
\end{proof}


\begin{corollary}[Expected reconstruction error]
\label{cor:init-global-error}
    For the recursive reconstruction above,
    \[
        \mathbb E\!\left[
            1-|\langle\psi|\widehat\psi_{\varnothing}\rangle|^2
        \right]
        \le
        4\sqrt C\sum_{\ell=0}^{n-1}
        \left(
            \frac1{\sqrt{N_\ell}}
            +
            \sqrt{\frac{2^\ell}{N_\ell S_\ell}}
        \right).
    \]
\end{corollary}
\begin{proof}
    For each level $\ell$, summing Lemma~\ref{lem:init-gluing} over $x\in\{0,1\}^{\ell}$ gives
    \begin{align*}
        &\sum_{x\in\{0,1\}^{\ell}}
        p_x\bigl(1-|\langle\psi_x|\widehat\psi_x\rangle|^2\bigr)\\
        &\le
        \sum_{x\in\{0,1\}^{\ell}}\sum_{b\in\{0,1\}}
        p_{xb}\bigl(
            1-|\langle\psi_{xb}|\widehat\psi_{xb}\rangle|^2
        \bigr)
        +2\sum_{x\in\{0,1\}^{\ell}}
        \|\widehat B_x-B_x\|_{\mathrm{op}}\\
        &=
        \sum_{y\in\{0,1\}^{\ell+1}}
        p_y\bigl(1-|\langle\psi_y|\widehat\psi_y\rangle|^2\bigr)
        +2\sum_{x\in\{0,1\}^{\ell}}
        \|\widehat B_x-B_x\|_{\mathrm{op}}.
    \end{align*}
    Summing over $\ell=0,\ldots,n-1$ gives a telescoping sum, in which the errors at levels $1,\ldots,n-1$ cancel. At the leaves ($\ell=n$), no qubits remain to be reconstructed, so their errors are zero. At the root ($\ell=0$), $p_{\varnothing}=1$ and $|\psi_{\varnothing}\rangle=|\psi\rangle$. We therefore obtain
    \[
        1-|\langle\psi|\widehat\psi_{\varnothing}\rangle|^2
        \le
        2\sum_{\ell=0}^{n-1}\sum_{x\in\{0,1\}^{\ell}}
        \|\widehat B_x-B_x\|_{\mathrm{op}}.
    \]
    Taking expectations and applying Lemmas~\ref{lem:cauchy-schwarz} and~\ref{lem:init-core-estimation}, together with $\sum_xp_x=1$, we obtain
    \begin{align*}
        \mathbb E\!\left[
            1-|\langle\psi|\widehat\psi_{\varnothing}\rangle|^2
        \right]
        &\le
        2\sum_{\ell=0}^{n-1}\sum_{x\in\{0,1\}^{\ell}}
        \mathbb E\|\widehat B_x-B_x\|_{\mathrm{op}}\\
        &\le
        2\sum_{\ell=0}^{n-1}\sum_{x\in\{0,1\}^{\ell}}
        \sqrt{\mathbb E\|\widehat B_x-B_x\|_F^2}
        \tag*{(by Lemma~\ref{lem:cauchy-schwarz})}
        \\
        &\le
        4\sqrt C\sum_{\ell=0}^{n-1}\frac1{\sqrt{N_\ell}}
        \sum_{x\in\{0,1\}^{\ell}}
        \sqrt{p_x^2+\frac{p_x}{S_\ell}}\\
        &\le
        4\sqrt C\sum_{\ell=0}^{n-1}
        \left(
            \frac1{\sqrt{N_\ell}}
            +
            \frac{\sum_{x\in\{0,1\}^{\ell}}\sqrt{p_x}}
                 {\sqrt{N_\ell S_\ell}}
        \right)\\
        &\le
        4\sqrt C\sum_{\ell=0}^{n-1}
        \left(
            \frac1{\sqrt{N_\ell}}
            +
            \sqrt{\frac{2^\ell}{N_\ell S_\ell}}
        \right).
        \tag*{\(\displaystyle
            \left(\text{by}~~\sum_{x\in\{0,1\}^{\ell}}\sqrt{p_x}
            \le 2^{\ell/2}\right)
        \)}
    \end{align*}
\end{proof}

We now combine the preceding bounds to prove Theorem~\ref{thm:initialization}.
    
\begin{proof}[Proof of Theorem~\ref{thm:initialization}]
    Let $K>0$ be a constant to be chosen later, and set
    \[
        N_\ell:=\lceil Kn^2\rceil,
        \qquad
        S_\ell:=
        \left\lceil
            \frac{Kd^{2/3}2^{\ell/3}}{N_\ell}
        \right\rceil,
        \qquad \ell=0,\ldots,n-1.
    \]
    Let $|\sigma_0\rangle:=|\widehat\psi_{\varnothing}\rangle$.
    By Corollary~\ref{cor:init-global-error},
    \begin{align*}
        \mathbb E[1-|\langle\psi|\sigma_0\rangle|^2]
        &\le
        4\sqrt{\frac CK}
        \left(
            1+d^{-1/3}\sum_{\ell=0}^{n-1}2^{\ell/3}
        \right)\\
        &\le
        4\sqrt{\frac CK}
        \left(1+\frac1{2^{1/3}-1}\right)
        \le\frac1{24},
    \end{align*}
    where the last inequality holds for a sufficiently large constant $K=O(C)$.
    Applying Lemma~\ref{lem:markov} gives
    \[
        \Pr\!\left[
            1-|\langle\psi|\sigma_0\rangle|^2>\frac12
        \right]
        \le
        2\,\mathbb E[1-|\langle\psi|\sigma_0\rangle|^2]
        \le\frac1{12}.
    \]
    Thus the required overlap holds with probability at least $\frac{11}{12}$, and the output is normalized by construction.
    
    The number of sampled bases is $\sum_{\ell=0}^{n-1}N_\ell=O(n^3)$ and the total copy count is
    \begin{align*}
        \sum_{\ell=0}^{n-1}N_\ell S_\ell
        =
        O\left(d^{2/3}\sum_{\ell=0}^{n-1}2^{\ell/3}\right)=O(d).
    \end{align*}
    All bases and shot counts are chosen before observing any outcomes, so the protocol is nonadaptive and single-copy. Its depth is $O(\log n)$ by the assumed properties of the suffix unitary ensembles.
\end{proof}

\section{Classical post--processing}
We assume that the experimental measurement results are provided as an outcome histogram for each measurement basis. Given these histograms, both initialization of $|\sigma_0\rangle$ and shadow bootstrapping can be performed in $\widetilde O(d)$ time and working memory. We first establish the computational primitives used by both procedures.

\begin{lemma}[Clifford circuit application]
\label{lem:clifford-vector-action}
    For a measurement circuit $U$ in our two-layer Clifford ensemble and any $x\in\mathbb C^d$, both $Ux$ and $U^\dagger x$ can be computed in $O(dn\log n)$ time and $O(d)$ working storage.
\end{lemma}
\begin{proof}
    We implement each $k$-qubit Clifford block by an $O(k)$-depth circuit over $\{H,S,\mathrm{CX}\}$. Each circuit layer contains at most $k$ gates, so each block uses $O(k^2)$ gates~\cite{maslov2007linear}. The two-layer ensemble has $n/k$ blocks per layer. With our choice $k=\Theta(\log n)$, the total gate count is
    \[
        \frac {2n}{k}\,O(k^2)
        =O(nk)
        =O(n\log n).
    \]
    
    For a dense vector $x\in\mathbb C^d$, an $H$ gate applies a fixed $2\times2$ transformation to each of $d/2$ amplitude pairs differing only in the target bit. An $S$ gate multiplies the $d/2$ amplitudes whose target bit is $1$ by $i$. A $\mathrm{CX}$ gate swaps $d/4$ amplitude pairs whose control bit is $1$ and whose target bits differ. Each gate therefore costs $O(d)$ arithmetic operations. Applying all $O(n\log n)$ gates costs $O(dn\log n)$. The same argument applies to the inverse circuit $U^{\dagger}$. All updates can be performed in place, so $O(d)$ working storage suffices.
\end{proof}

\begin{lemma}[Exact visibility evaluation]
\label{lem:efficient-visibility}
    For a measurement circuit $U$ in our two-layer Clifford ensemble and a Pauli operator $P$, the visibility $m_{U^\dagger P U}$ can be computed in $O(n\log n)$ time and $O(n)$ working storage.
\end{lemma}
\begin{proof}
    Our two-layer Clifford circuit contains $O(n\log n)$ gates
    from $\{H,S,\mathrm{CX}\}$. Conjugation by $H$ or $S$ updates one Pauli factor, while conjugation by $\mathrm{CX}$ updates at most two. Each update therefore takes $O(1)$ time, and the same holds for the adjoint gates.
    Thus, the Pauli label of $U^\dagger P U$ can be computed in $O(n\log n)$ time and $O(n)$ working storage. We ignore the overall sign since it does not affect the visibility.
    
    Next, by Lemma~\ref{lem:visibility-transfer}, the support of a Pauli string on the blocks of the first layer can be determined in $O(n)$ time.
    Its visibility is then obtained by multiplying $n/k$ matrices of size $2\times2$, requiring $O(n/k)$ arithmetic operations. Together with the Pauli conjugation step, this gives $O(n\log n)$ time and $O(n)$ working memory.
\end{proof}

\begin{lemma}[Shadow action]
\label{lem:weighted-shadow-action}
    Let $\mathcal M$ be the  shadow channel of our two-layer Clifford ensemble. Given a measurement circuit $U$ and vectors $h=(h_z)_z,x\in\mathbb C^d$, the vector
    \[
        \left[
            \sum_z h_z\,
            \mathcal M^{-1}
            \left(U^\dagger|z\rangle\langle z|U\right)
        \right]x
        \in\mathbb C^d
    \]
    can be computed in $\widetilde{O}(d)$ time and $\widetilde{O}(d)$ working memory.
\end{lemma}
\begin{proof}
    For $a\in\{0,1\}^n$, write $Z^a:=\bigotimes_{j=1}^n Z^{a_j}$ and $\mathcal Z:=\{Z^a:a\in\{0,1\}^n\}$. Since $Z^a|z\rangle=(-1)^{a\cdot z}|z\rangle$, for any $v\in\mathbb C^d$,
    \[
        \sum_a v_a Z^a
        =
        \sum_z\left(\sum_a(-1)^{a\cdot z}v_a\right)
        |z\rangle\langle z|
        =
        \texttt{diag}\!\left(\mathrm{WH}^{-1}(v)\right),
    \]
    where $\texttt{diag}(v)$ denotes the diagonal matrix whose diagonal entries are the components of $v$.
    
    Using the Pauli expansion and $\mathcal M^{-1}(P)=m_P^{-1}P$, we obtain
    \begin{align*}
        \sum_z h_z\,
        \mathcal M^{-1}\!\left(U^\dagger|z\rangle\langle z|U\right)
        &=
        \frac1d\sum_z\sum_P
        h_zm_P^{-1}
        \operatorname{Tr}\!\left(
            UPU^\dagger|z\rangle\langle z|
        \right)P\\
        &=
        \frac1d\sum_z
        \sum_{P:\,UPU^\dagger\in\pm\mathcal Z}
        h_zm_P^{-1}
        \operatorname{Tr}\!\left(
            UPU^\dagger|z\rangle\langle z|
        \right)P\\
        &=
        \frac1d\sum_z\sum_a
        h_zm_{U^\dagger Z^aU}^{-1}
        (-1)^{a\cdot z}U^\dagger Z^aU\\
        &=
        U^\dagger\left(
            \sum_a m_{U^\dagger Z^aU}^{-1}
            \left[\frac1d\sum_z(-1)^{a\cdot z}h_z\right]Z^a
        \right)U\\
        &=
        U^\dagger\left(
            \sum_a m_{U^\dagger Z^aU}^{-1}
            \mathrm{WH}(h)_a Z^a
        \right)U\\
        &=
        U^\dagger\left(\sum_a\mu_a Z^a\right)U\\
        &=
        U^\dagger
        \texttt{diag}\!\left(\mathrm{WH}^{-1}(\mu)\right)U,
    \end{align*}
    where $\mu_a:=m_{U^\dagger Z^aU}^{-1}\mathrm{WH}(h)_a$. Here $P$ ranges over unsigned Pauli operators, and we use $m_{-P}=m_P$. When $UPU^\dagger=\pm Z^a$, the signs in the trace coefficient and in $P=\pm U^\dagger Z^aU$ cancel.
    
    First, we compute $\mathrm{WH}^{-1}(\mu)$. Reading $h$ and performing the two fast Walsh--Hadamard transforms costs $O(d\log d)$ time. The $d$ visibility evaluations take $\widetilde{O}(d)$ total time by Lemma~\ref{lem:efficient-visibility}. These computations use $\widetilde{O}(d)$ working memory. The required vector is
    \begin{equation}
        U^\dagger
        \texttt{diag}\!\left(\mathrm{WH}^{-1}(\mu)\right)Ux.
    \end{equation}
    By Lemma~\ref{lem:clifford-vector-action}, computing $Ux$ and applying $U^\dagger$ costs $O(dn\log n)$ time and $O(d)$ storage. Multiplication by the diagonal matrix uses only its $d$ diagonal entries and takes $O(d)$ time. The total time and working memory are therefore $\widetilde O(d)$.
\end{proof}

\subsection{Initialization}

For each $x\in\{0,1\}^{\ell}$, we reconstruct
\[
    |\widehat\psi_x\rangle
    =
    \widehat\alpha_0\,|0\rangle\otimes|\widehat\psi_{x0}\rangle
    +
    \widehat\alpha_1\,|1\rangle\otimes|\widehat\psi_{x1}\rangle.
\]
The coefficients are obtained by estimating the
$2\times2$ matrix $\widehat B_x$ and computing
a normalized eigenvector corresponding to its largest eigenvalue.

Each histogram has at most $d$ entries, so grouping them
by prefix $x$ takes $O(d)$ time. There are $O(n^3)$ histograms, giving $\widetilde O(d)$ total time.
Applying the construction of Lemma~\ref{lem:weighted-shadow-action} to the inverse shadow channel $\mathcal M_\ell^{-1}$ on the suffix, followed by inner products with $|q_0\rangle$ and $|q_1\rangle$, estimates each entry of $B_x$ in $\widetilde O(N_\ell 2^{n-\ell})$ time.
Since $N_\ell=O(n^2)$, estimating and diagonalizing $\widehat B_x$ and constructing $|\widehat\psi_x\rangle$ take $\widetilde O(2^{n-\ell})$ time per node. There are $2^\ell$ nodes at level $\ell$, so each level costs $\widetilde O(2^n)$ time. Repeating over $\ell=n-1,\ldots,0$ therefore takes $\widetilde O(2^n)=\widetilde O(d)$ total time, including histogram processing.

The state vectors at each level have total length $2^\ell 2^{n-\ell}=d$. After reconstructing all states at level $\ell$, we delete the estimates at level $\ell+1$ from memory. Thus, only two adjacent levels need to be stored at a time, and $\widetilde O(d)$ working memory suffices.

\subsection{Shadow Bootstrapping}

At stage $t$, we compute
\begin{equation}
    |\widetilde\sigma_{t}\rangle
    =
    \left(
        \sigma_{t-1}+\frac1{N_t}
        \sum_{i=1}^{N_t}\Delta_{t,i}
    \right)|\sigma_0\rangle.
\end{equation}
For each basis $U_{t,i}$, computing the predicted probabilities $p_z:=\operatorname{Tr}(U_{t,i}\sigma_{t-1} U_{t,i}^\dagger |z\rangle\langle z|)$ for all $z$ takes $\widetilde O(d)$ time by Lemma~\ref{lem:clifford-vector-action}. We then compute the weights $h_z:=\widehat f_{t,i}(z)-p_z$ in $O(d)$ additional time. 
Lemma~\ref{lem:weighted-shadow-action} then computes $\Delta_{t,i}|\sigma_0\rangle$ in $\widetilde O(d)$ time. Computing $\sigma_{t-1}|\sigma_0\rangle= |\sigma_{t-1}\rangle \langle \sigma_{t-1}| \sigma_0 \rangle$ takes $O(d)$ time.

The choice $N_t=O(1)$ in the proof of Corollary~\ref{cor:bootstrap-convergence} therefore gives $\widetilde O(N_t d)=\widetilde O(d)$ time per stage. Since the same corollary uses $J=O(\log(1/\epsilon))$ stages, the total time is $\widetilde O(Jd)=\widetilde O(d)$.

We process the measurement bases one at a time and reuse the same memory for each basis. Only a constant number of length-$d$ vectors need to be stored, so the working memory is $\widetilde O(d)$.

\section{Full Algorithm}
\begin{algorithm}[H]
\caption{Nonadaptive measurement schedule}
\label{alg:measurement-schedule}
\begin{algorithmic}[1]
\Input Copy access to an unknown $n$-qubit pure state
       $|\psi\rangle$; target infidelity $\epsilon\in(0,1)$.
\Output Initialization data $\mathcal D_{\mathrm{init}}$
        and refinement data $\mathcal D_{\mathrm{ref}}$,
        including circuit descriptions, shot counts, and per-basis outcome histograms.

\State $d\gets2^n$.
\State Choose sufficiently large constant $K$ as in the proof of
       Theorem~\ref{thm:initialization}.
\State $J\gets\lceil\log_2(6/(\epsilon))\rceil$
       \Comment{Corollary~\ref{cor:bootstrap-success}}

\For{$\ell=0,\ldots,n-1$}
    \State $N_\ell\gets\lceil Kn^2\rceil$, $S_\ell\gets
           \left\lceil Kd^{2/3}2^{\ell/3}/N_\ell\right\rceil$
           \Comment{Theorem~\ref{thm:initialization}}
    \For{$i=1,\ldots,N_\ell$}
        \State Independently sample
               $U_{\ell,i}\sim\mathcal E_\ell$.
        \State Schedule $S_\ell$ shots using
               $I_{2^\ell}\otimes U_{\ell,i}$.
    \EndFor
\EndFor

\For{$t=1,\ldots,J$}
    \State $N_t\gets\lceil16(C+1)\rceil$, $S_t\gets d\,2^{t-1}$
           \Comment{Corollary~\ref{cor:bootstrap-convergence}}
    \For{$i=1,\ldots,N_t$}
        \State Independently sample $U_{t,i}\sim\mathcal E$.
        \State Schedule $S_t$ shots using $U_{t,i}$.
    \EndFor
\EndFor

\State Execute all scheduled measurements on disjoint copies,
       with computational-basis readout after each circuit.
\State Store the initialization circuits, shot counts, and
       histograms $\widehat f_{\ell,i}(x,b)$ in
       $\mathcal D_{\mathrm{init}}$.
\State Store the refinement circuits, shot counts, and
       histograms $\widehat f_{t,i}(z)$ in
       $\mathcal D_{\mathrm{ref}}$.
\State \Return
       $\mathcal D_{\mathrm{init}},\mathcal D_{\mathrm{ref}}$.
\end{algorithmic}
\end{algorithm}

\begin{algorithm}[H]
\caption{Weighted shadow action}
\label{alg:shadow-action}
\begin{algorithmic}[1]
\Input A measurement circuit $U$; weights $h=(h_z)_z$;
       a vector $x$ on the same register;
       the matched shadow channel $\mathcal M$.
\Output
$\left[\sum_z h_z\mathcal M^{-1}
(U^\dagger|z\rangle\langle z|U)\right]x$
(Lemma~\ref{lem:weighted-shadow-action}).

\Function{ShadowAction}{$U,h,x,\mathcal M$}
    \State $\mu\gets\mathrm{WH}(h)$.
    \For{each bit string $a$ on the register of $U$}
        \State $\mu_a\gets\mu_a/m_{U^\dagger Z^aU}$
               \Comment{Lemma~\ref{lem:efficient-visibility}}
    \EndFor
    \State $\lambda\gets\mathrm{WH}^{-1}(\mu)$.
    \State $y\gets Ux$.
    \State $y_z\gets\lambda_z y_z$ for every $z$.
    \State \Return $U^\dagger y$.
\EndFunction
\end{algorithmic}
\end{algorithm}

\begin{algorithm}[H]
\caption{Initialization}
\label{alg:initialization}
\begin{algorithmic}[1]
\Input Initialization data $\mathcal D_{\mathrm{init}}$
       from Algorithm~\ref{alg:measurement-schedule};
       suffix shadow channels
       $\{\mathcal M_\ell\}_{\ell=0}^{n-1}$ (Lemma~\ref{lem:suffix-ensembles}).
\Output A normalized initial estimate $|\sigma_0\rangle$
        (Theorem~\ref{thm:initialization}).

\Function{Initialization}{
    $\mathcal D_{\mathrm{init}},
    \{\mathcal M_\ell\}_{\ell=0}^{n-1}$}
    \State Read $n$, $N_\ell$, $S_\ell$, the circuits,
           and the histograms from $\mathcal D_{\mathrm{init}}$.
    \State $|\widehat\psi_x\rangle\gets1\in\mathbb C$
           for every leaf $x\in\{0,1\}^n$.

    \For{$\ell=n-1,\ldots,0$}
        \For{$x\in\{0,1\}^{\ell}$}
            \State $|q_0\rangle\gets
                   |0\rangle\otimes|\widehat\psi_{x0}\rangle$,
                   $|q_1\rangle\gets
                   |1\rangle\otimes|\widehat\psi_{x1}\rangle$.
            \State $h_i\gets
                   (\widehat f_{\ell,i}(x,b))_{b\in\{0,1\}^{n-\ell}}$
                   for $i=1,\ldots,N_\ell$.
            \State $(\widehat B_x)_{rs}\gets
                   \dfrac{1}{N_\ell}\sum_{i=1}^{N_\ell}
                   \langle q_r|
                   \mathsf{ShadowAction}
                   (U_{\ell,i},h_i,|q_s\rangle,\mathcal M_\ell)$
                   for $r,s\in\{0,1\}$.
            \State Choose any normalized eigenvector $\widehat\alpha$
                   corresponding to the largest eigenvalue of $\widehat B_x$.
            \State $|\widehat\psi_x\rangle\gets
                   \widehat\alpha_0|q_0\rangle+
                   \widehat\alpha_1|q_1\rangle$.
        \EndFor
    \EndFor

    \State \Return
           $|\sigma_0\rangle
           \gets|\widehat\psi_{\varnothing}\rangle$.
\EndFunction
\end{algorithmic}
\end{algorithm}

\begin{algorithm}[H]
    \caption{Shadow bootstrapping}
    \label{alg:shadow-bootstrapping}
    \begin{algorithmic}[1]
    \Input Data $\mathcal D_{\mathrm{init}},
           \mathcal D_{\mathrm{ref}}$
           from Algorithm~\ref{alg:measurement-schedule};
           shadow channels
           $\{\mathcal M_\ell\}_{\ell=0}^{n-1}$
           and $\mathcal M$.
    \Output A normalized estimate $|\widehat\psi\rangle$.
    
    \State Read $J$, $N_t$, $S_t$, the circuits,
           and the histograms from $\mathcal D_{\mathrm{ref}}$.
    \State $|\sigma_0\rangle\gets$
    \Call{Initialization}{
        $\mathcal D_{\mathrm{init}},
        \{\mathcal M_\ell\}_{\ell=0}^{n-1}$}.
    
    \For{$t=1,\ldots,J$}
        \State $|\widetilde\sigma_t\rangle\gets
               \langle\sigma_{t-1}|\sigma_0\rangle
               |\sigma_{t-1}\rangle$.
    
        \For{$i=1,\ldots,N_t$}
            \State $h_z\gets\widehat f_{t,i}(z)
                   -\operatorname{Tr}\!\left(
                       U_{t,i}\sigma_{t-1}U_{t,i}^\dagger
                       |z\rangle\langle z|
                   \right)$
                   for every $z\in\{0,1\}^n$.
            \State $|\widetilde\sigma_t\rangle\gets
                   |\widetilde\sigma_t\rangle
                   +\dfrac1{N_t}$
            \Call{ShadowAction}{\!
                $U_{t,i},h,|\sigma_0\rangle,\mathcal M$\,}.
        \EndFor
    
        \If{$\|\widetilde\sigma_t\|_2>0$}
            \State $|\sigma_t\rangle\gets
                   |\widetilde\sigma_t\rangle/
                   \|\widetilde\sigma_t\|_2$.
        \Else
            \State $|\sigma_t\rangle\gets|\sigma_{t-1}\rangle$.
        \EndIf
    \EndFor
    
    \State \Return
           $|\widehat\psi\rangle\gets|\sigma_J\rangle$.
    \end{algorithmic}
\end{algorithm}

\begin{theorem}[Full algorithm at constant confidence]
\label{thm:full-constant} 
    For any $n$-qubit pure state $|\psi\rangle$ and $\epsilon\in(0,1)$, Algorithms~\ref{alg:measurement-schedule} and~\ref{alg:shadow-bootstrapping} output a unit vector $|\widehat\psi\rangle$ satisfying
    \[
        \Pr\!\left[
            1-|\langle\psi|\widehat\psi\rangle|^2
            \le\epsilon
        \right]\ge\frac56.
    \]
    The protocol uses nonadaptive single-copy Clifford measurements of depth $O(\log n)$, with
    \[
        T=O(d/\epsilon),
        \qquad
        N=O\!\left(n^3+\log(2/\epsilon)\right).
    \]
    Given per-basis outcome histograms, classical post-processing takes $\widetilde O(d\log(2/\epsilon))$ time and $\widetilde O(d)$ working memory.
\end{theorem}

\begin{proof}
    By Theorem~\ref{thm:initialization}, initialization outputs a normalized vector $|\sigma_0\rangle$ with $|\langle\psi|\sigma_0\rangle|^2\ge1/2$ except with probability at most $1/12$. Conditional on any such initial estimate, the refinement data remain independent of initialization. Corollary~\ref{cor:bootstrap-success} therefore gives
    \[
        \Pr\!\left[
            1-|\langle\psi|\widehat\psi\rangle|^2>\epsilon
        \right]
        \le\frac1{12}+\frac1{12}
        =\frac16.
    \]

    Theorem~\ref{thm:initialization} and
    Corollary~\ref{cor:bootstrap-convergence} give
    \begin{align*}
        T
        &=
        \sum_{\ell=0}^{n-1}N_\ell S_\ell
        +
        \sum_{t=1}^{J}N_tS_t
        =
        O(d)+O(d/\epsilon)
        =
        O(d/\epsilon),\\
        N
        &\le
        \sum_{\ell=0}^{n-1}N_\ell
        +
        \sum_{t=1}^{J}N_t
        =
        O\!\left(n^3+\log(1/\epsilon)\right).
    \end{align*}
    All bases and shot counts are chosen before observing any outcomes, and each measurement acts on one copy. The measurement depth is $O(\log n)$ by the construction of the measurement ensembles.

    Lemma~\ref{lem:weighted-shadow-action} ensures that the pseudocode computes the initialization matrices and residual updates analyzed above. The preceding classical post-processing analysis gives the stated time and working-memory bounds, including histogram processing, normalization, and state-vector output. 
\end{proof}

We use the following standard confidence-amplification
lemma.

\begin{lemma}[Confidence amplification
    {\cite[Proposition~2.4]{haah2023query}}]
\label{lem:confidence-amplification}
    Let $\mathcal A$ estimate an unknown point $x_\star$ in a metric space $(\mathcal X,D)$ with efficiently computable distances. Suppose its output $X$ satisfies $\Pr[D(X,x_\star)\le r]\ge2/3$.
    For any $0<\delta<1/2$, running $\mathcal A$ independently $O(\log(1/\delta))$ times and selecting one output using pairwise distances yields an estimate $\widehat X$ satisfying
    \[
        \Pr[D(\widehat X,x_\star)\le3r]\ge1-\delta.
    \]
\end{lemma}

\begin{theorem}[Arbitrary confidence]
\label{thm:full-confidence}
    For any $\epsilon\in(0,1)$ and $\delta\in(0,1/2)$, there is a nonadaptive single-copy tomography algorithm that outputs a unit vector $|\widehat\psi\rangle$ satisfying
    \[
        \Pr\!\left[
            1-|\langle\psi|\widehat\psi\rangle|^2
            \le\epsilon
        \right]\ge1-\delta.
    \]
    Each measurement has Clifford depth $O(\log n)$, and
    \[
        T=
        O\!\left(
            \frac d\epsilon\log\frac1\delta
        \right),
        \qquad
        N=
        O\!\left(
            \left[n^3+\log\frac1\epsilon\right]
            \log\frac1\delta
        \right).
    \]
    Given per-basis outcome histograms, classical
    post-processing takes
    \[
        \widetilde O\!\left(
            d\left[
                \log\frac1\epsilon\log\frac1\delta
                +
                \log^2\frac1\delta
            \right]
        \right)
    \]
    time and $\widetilde O(d\log(1/\delta))$
    working memory.
\end{theorem}

\begin{proof}
    Run the entire algorithm of Theorem~\ref{thm:full-constant}, including initialization, independently $R=O(\log(1/\delta))$ times with target infidelity $\epsilon/9$.
    In Lemma~\ref{lem:confidence-amplification}, take $D$ to be the trace distance between pure states:
    \[
        D(\phi,\psi)
        :=
        \frac12
        \bigl\|
            |\phi\rangle\langle\phi|
            -
            |\psi\rangle\langle\psi|
        \bigr\|_{\mathrm{tr}}
        =
        \sqrt{1-|\langle\phi|\psi\rangle|^2}.
    \]
    By Eq.~\eqref{eq:prelim-pure-distances}, each output has distance at most $\sqrt{\epsilon}/3$ from $\psi$ with probability at least $5/6$. The lemma therefore selects an output with trace distance at most $\sqrt{\epsilon}$, and hence infidelity at most $\epsilon$, with probability at least $1-\delta$.
    
    The bounds on sample complexity $T$ and the number of sampled measurement bases $N$ follow by multiplying the corresponding bounds in Theorem~\ref{thm:full-constant} by $R$. All measurements can be scheduled in advance, preserving nonadaptivity and depth.
    Each pairwise distance costs $O(d)$ time to compute. Candidate selection thus adds $O(dR^2)$ time and $O(dR)$ memory, with no additional measurements. Together with the $R$ executions, this gives the stated computational bounds. This also proves Theorem~\ref{thm:main}.
\end{proof}

\section*{Acknowledgments}
This work was supported by grants from the National Research Foundation of Korea (NRF) funded by the Korean Government [Ministry of Science and ICT (MSIT)] (Grants RS-2023-00283291, RS-2023-NR077112, RS-2024-00413957, and Quantum Technology R\&D Leading Program (Quantum Computing) (Grant RS-2024-00442994)), and a core center program grant funded by the Ministry of Education (No.~2021R1A6C101B418). This research was conducted as part of an R\&D project supported by SDT Inc.\ and SAIT.

\section*{AI Disclosure}
GPT-5.6 and GPT-6 were used throughout the research to discuss technical ideas, examine proofs, analyze prior work, and assist with manuscript preparation. The authors alternated between discussions with these models and their own checks and revisions of the arguments. These exchanges helped clarify and simplify parts of the analysis, including coefficient recovery using the reconstruction matrix $B_x$ and the allocation of measurement bases and shots during initialization. They also contributed to the runtime analysis of computing $Ux$ for shallow circuits and histogram-based post-processing with the inverse shadow channel. The authors reviewed the resulting material and take responsibility for the mathematical claims, references, and writing in the final manuscript.

\bibliographystyle{alpha}
\bibliography{references}

\appendix
\section{Deferred Proofs}
\shadowsuminter*

\begin{proof}
    Let $P$ range over $\{I,X,Y,Z\}^{\otimes n}$,
    and let $\mathcal Z:=\{I,Z\}^{\otimes n}$.
    Since $U$ is Clifford,
    \[
        \sum_b
        \operatorname{Tr}\!\left(
            PU^\dagger|b\rangle\langle b|U
        \right)
        U^\dagger|b\rangle\langle b|U
        =
        \begin{cases}
            P, & UPU^\dagger\in\pm\mathcal Z,\\
            0, & UPU^\dagger\notin\pm\mathcal Z.
        \end{cases}
    \]
    Using $\mathcal M^{-1}(P)=m_P^{-1}P$ in Eq.~\eqref{eq:inver_shadow_P}, we obtain
    \begin{align*}
        &\sum_b
        \operatorname{Tr}(AU^\dagger|b\rangle\langle b|U)
        \mathcal M^{-1}(U^\dagger|b\rangle\langle b|U)
        \\
        &=
        \frac1d\sum_b\sum_P m_P^{-1}
        \operatorname{Tr}(AU^\dagger|b\rangle\langle b|U)
        \operatorname{Tr}(PU^\dagger|b\rangle\langle b|U)P
        \\
        &=
        \frac1d\sum_P m_P^{-1}
        \operatorname{Tr}\!\left[
            A\sum_b
            \operatorname{Tr}(PU^\dagger|b\rangle\langle b|U)
            U^\dagger|b\rangle\langle b|U
        \right]P\\
        &=
        \frac1d
        \sum_{P:UPU^{\dagger}\in \pm \mathcal{Z}}
        m_P^{-1}\operatorname{Tr}(AP)P\\
        &=
        \frac1d\sum_b\sum_P m_P^{-1}
        \operatorname{Tr}(AP)
        \operatorname{Tr}(PU^\dagger|b\rangle\langle b|U)
        U^\dagger|b\rangle\langle b|U\\
        &=
        \sum_b
        \operatorname{Tr}\!\left[
            A\left(
                \frac1d\sum_P m_P^{-1}
                \operatorname{Tr}(PU^\dagger|b\rangle\langle b|U)P
            \right)
        \right]
        U^\dagger|b\rangle\langle b|U
        \\
        &=
        \sum_b
        \operatorname{Tr}\!\left(
            A\mathcal M^{-1}(U^\dagger|b\rangle\langle b|U)
        \right)
        U^\dagger|b\rangle\langle b|U.
    \end{align*}
\end{proof}

\initcomplexvariance*

\begin{proof}
    Write $O=O_R+iO_I$, where $O_R=(O+O^\dagger)/2$ and $O_I=(O-O^\dagger)/(2i)$ are Hermitian. By linearity of the estimator and Lemma~\ref{lem:init-core-variance}, $\mathbb E[\widehat o(O)]=p_x\operatorname{Tr}(O\psi_x)$.
    Moreover,
    \begin{align*}
        \mathbb E\!\left[
            \left|\widehat o(O)-p_x\operatorname{Tr}(O\psi_x)\right|^2
        \right]
        &=
        \mathbb E\!\left[
            \left(
                \widehat o(O_R)
                -p_x\operatorname{Tr}(O_R\psi_x)
            \right)^2
        \right]+
        \mathbb E\!\left[
            \left(
                \widehat o(O_I)
                -p_x\operatorname{Tr}(O_I\psi_x)
            \right)^2
        \right]\\
        \tag*{(by Lemma~\ref{lem:init-core-variance})}
        &\le
        \frac{C\bigl(
            \|O_R\|_F^2+
            \|O_I\|_F^2
        \bigr)}{N_\ell}
        \left(p_x^2+\frac{p_x}{S_\ell}\right)\\
        &=
        \frac{C\|O\|_F^2}{N_\ell}
        \left(p_x^2+\frac{p_x}{S_\ell}\right).
    \end{align*}
\end{proof}

\begin{lemma}[{Extension from~\cite[Lemma~25]{cho2026quantum}}]
\label{lem:tensor-shadow-bound}
    Let $\mathcal E_A$ and $\mathcal E_B$ be independent
    unitary ensembles on $n_A$ and $n_B$ qubits,
    respectively, with invertible shadow channels.
    Suppose that, for every state $\rho_A$ and Hermitian
    observable $O_A$ on $A$, and every state $\rho_B$
    and Hermitian observable $O_B$ on $B$, the matched
    shadow estimators satisfy
    \[
        \operatorname{Var}
        \left[\operatorname{Tr}(O_A\widehat\rho_A)\right]
        \le C_A\|O_A\|_{F}^2,
        \qquad
        \operatorname{Var}
        \left[\operatorname{Tr}(O_B\widehat\rho_B)\right]
        \le C_B\|O_B\|_{F}^2.
    \]
    Then, for every joint state $\rho_{AB}$ and Hermitian
    observable $O_{AB}$, the matched shadow estimator
    for $\mathcal E_A\otimes\mathcal E_B$ satisfies
    \[
        \operatorname{Var}
        \left[
            \operatorname{Tr}(O_{AB}\widehat\rho_{AB})
        \right]
        \le (C_A+1)(C_B+1)\|O_{AB}\|_{F}^2.
    \]
\end{lemma}
\begin{proof}
    By unbiasedness and $\|\rho_A\|_{F}\le1$, the assumed
    variance bound implies
    \[
        \mathbb E
        \left[\operatorname{Tr}(O_A\widehat\rho_A)^2\right]
        \le (C_A+1)\|O_A\|_{F}^2.
    \]
    The analogous bound on $B$ holds with constant $C_B+1$. Write
    \[
        O=\sum_P P\otimes X_P,
    \]
    where $P$ ranges over the Pauli operators on $A$,
    and each $X_P=2^{-n_A}\operatorname{Tr}_A(O(P\otimes I))$ is a Hermitian operator on $B$.
    Pauli orthogonality gives
    \[
        \|O\|_{F}^2=2^{n_A}\sum_P\|X_P\|_{F}^2.
    \]
    
    The shadow channel factorizes $\mathcal{M}_{AB}=\mathcal{M}_A \otimes \mathcal{M}_B$, so
    $\widehat\rho_{AB}
    =\widehat\rho_A\otimes\widehat\rho_B$. Then
    \begin{align*}
        \operatorname{Tr}(O\widehat \rho_{AB}) 
        &= 
        \sum_P \operatorname{Tr}(P\widehat \rho_{A})\operatorname{Tr}(X_P\widehat \rho_{B})\\
        &=
        \sum_P \operatorname{Tr}(P\widehat \rho_{A}) \alpha_P\\
        &=
        \operatorname{Tr}\left(\widetilde O_A \widehat\rho_A\right),
    \end{align*}
    where $\alpha_P:=\operatorname{Tr}(X_P\widehat\rho_B)$ and $\widetilde O_A:=\sum_P\alpha_P P$ are auxiliary observables on $A$.
    By Pauli orthogonality,
    \begin{equation}
        \|\widetilde O_A\|_F^2
        =
        2^{n_A}\sum_P\alpha_P^2
        =
        2^{n_A}\sum_P
        \operatorname{Tr}(X_P\widehat\rho_B)^2.
    \end{equation}
    The conditional state on $A$ is independent of $U_A$. We can therefore apply the assumed second-moment bound on $A$ with $(U_B,z_B)$ held fixed.
    \begin{align*}
        \mathbb E_{A,B}
        \left[\operatorname{Tr}(O\widehat\rho_{AB})^2\right]
        &=
        \mathbb E_B
        \left[
            \mathbb E_A
            \left[
                \operatorname{Tr}
                (\widetilde O_A\widehat\rho_A)^2
                \,\middle|\,U_B,z_B
            \right]
        \right]\\
        &\le
        (C_A+1)\,
        \mathbb E_B\left[\|\widetilde O_A\|_{F}^2\right]\\
        &=
        2^{n_A}(C_A+1)\sum_P
        \mathbb E_B
        \left[\operatorname{Tr}(X_P\widehat\rho_B)^2\right]\\
        &\le
        2^{n_A}(C_A+1)(C_B+1)\sum_P\|X_P\|_{F}^2\\
        &=
        (C_A+1)(C_B+1)\|O\|_{F}^2.
    \end{align*}
    Since variance is bounded by the second moment, the claim follows.
\end{proof}

\begin{lemma}[Shallow ensembles for all suffix sizes]
\label{appx:lem:suffix-ensembles}
    Let 
    $k:=\max\{8,2\lceil\log_2 n\rceil\}$.
    For each $\ell=0,\ldots,n-1$, write $n-\ell=mk+r$, where $m=\left\lfloor \frac{n-\ell}{k} \right\rfloor$ and $0 \le r \le k-1$. Define unitary ensemble
    \[
        \mathcal E_\ell
        :=
        \mathcal E_{\mathrm{2L}}
        \left(
            km
        \right)
        \otimes
        \mathrm{Cl}(r),
    \]
    where $\mathcal E_{\mathrm{2L}}$ is the cyclic two-layer Clifford ensemble with block size $k$.
    Let $\mathcal M_\ell$ be its shadow channel. For every $(n-\ell)$-qubit suffix state $\rho$ and Hermitian observable $O\in \mathcal{L}(H^{2^{n-\ell}})$,
    \[
        \operatorname{Var}
        \left[
            \operatorname{Tr}(O\widehat\rho)
        \right]
        \le C_\ell\|O\|_F^2,
        \qquad C_\ell=O(1),
    \]
    where
    $\widehat\rho
    :=\mathcal M_\ell^{-1}
    (U^\dagger|z\rangle\langle z|U)$,
    $U\sim\mathcal E_\ell$, and $z\in\{0,1\}^{n-\ell}$ is a measurement outcome.
\end{lemma}

\begin{proof}
    Factors acting on zero qubits are omitted.
    If $m=0$, the ensemble is simply $\mathrm{Cl}(r)$, whose variance constant is at most $3$~\cite{huang2020predicting}. For $m\ge1$, our choice of $k$ gives
    \[
        \frac{m}{2^{k/2}}
        \le
        \frac{n}{k2^{k/2}}
        \le \frac1k \le1.
    \]
    The variance bound for the two-layer Clifford ensemble therefore applies to $\mathcal E_{\mathrm{2L}}(km)$, with coefficient $8e^{45m/2^{k/2}}=O(1)$ multiplying $\|O\|_F^2$ (see~\cite[Theorem~7]{cho2026quantum} for the derivation). 
    The global Clifford factor $\mathrm{Cl}(r)$, when present, has variance constant at most $3$.
    Applying Lemma~\ref{lem:tensor-shadow-bound}, we obtain
    \[
        \operatorname{Var}
        \left[\operatorname{Tr}(O\widehat\rho)\right]
        \le
        (3+1)\left(8e^{45m/2^{k/2}}+1\right)\|O\|_F^2.
    \]
    Thus we may take $C_\ell:=4(8e^{45n/(2^{k/2}k)}+1)=O(1)$.
    When $r=0$, the same bound follows directly from the two-layer estimate.
    The matched shadow channel is invertible because the component channels are invertible. Finally, each Clifford block in the two-layer Clifford circuit has depth $O(k)$.
    Since $r<k$, the Clifford circuit from Cl($r$) also has depth at most $O(k)$.
    Hence the total depth is $O(k)=O(\log(n))$.
\end{proof}

\section{Low-Rank Tomography with Nearly Minimal Measurement Settings}
\label{app:sec:low-rank}

The shadow bootstrapping method developed in the main text also extends to nonadaptive single-copy tomography of low-rank quantum states. For states of rank at most $r$, our protocol uses $\widetilde O(r)$ measurement settings and $\widetilde O(dr^2/\epsilon^2)$ copies to achieve trace-norm error $\epsilon$ with failure probability at most $\delta$. 
The number of sampled measurement bases matches the $\Omega(r)$ lower bound up to polylogarithmic factors~\cite{heinosaari2013quantum, brandao2020fast}. The sample complexity likewise matches the $\Omega(dr^2/\epsilon^2)$ lower bound for nonadaptive single-copy tomography up to polylogarithmic factors~\cite{Lowe2025lower}.

\subsection{Preliminaries}

\begin{lemma}[{Matrix Bernstein inequality with spectral clipping~\cite[Lemma~4]{cho2026quantum}}]
\label{lem:clipped-matrix-bernstein}
    Let $\widehat X_1,\ldots,\widehat X_N$ be independent copies of a Hermitian random matrix $\widehat X\in\mathbb C^{d\times d}$ satisfying $\mathbb E[\widehat X]=X$ and $\|\mathbb E[\widehat X^2]\|_{\mathrm{op}}\le V$. For $0<\delta<1$, set $K:=\sqrt{VN/\log(2d/\delta)}$. Then, with probability at least $1-\delta$,
    \[
        \left\|
            \frac1N\sum_{i=1}^N
            \operatorname{clip}_K(\widehat X_i)-X
        \right\|_{\mathrm{op}}
        \le
        C_0\sqrt{\frac{V\log(2d/\delta)}{N}},
    \]
    where $C_0>0$ is a universal constant and $\operatorname{clip}_K(x)=\max\{-K,\min\{x,K\}\}$.
\end{lemma}

\begin{lemma}[Rank-$r$ spectral truncation]
\label{lem:rank-r-truncation}
    Let $A,B\in\mathbb C^{d\times d}$ be Hermitian, with $\operatorname{rank}(B)\le r$. Let $\mathrm{proj}_{r}(A) := \mathrm{argmin}_{\mathrm{rank}(X)\leq r}\|A-X\|_{\mathrm{op}}$ retain the $r$ eigenvalues of $A$ with largest absolute value and their corresponding eigenvectors. Then
    \[
        \|\operatorname{proj}_r(A)-B\|_F^2
        \le
        8r\|A-B\|_{\mathrm{op}}^2.
    \]
    \end{lemma}
\begin{proof}
    \begin{align*}
        \|\operatorname{proj}_r(A)-B\|_F^2 
        &\le
        \mathrm{rank}(\operatorname{proj}_r(A)-B) \|\operatorname{proj}_r(A)-B\|_{\mathrm{op}}^2\\
        &\le
        2r\|\operatorname{proj}_r(A)-B\|_{\mathrm{op}}^2\\
        &\le
        2r(\|\operatorname{proj}_r(A)-A\|_{\mathrm{op}}+\|A-B\|_{\mathrm{op}})^2\\
        &\le
        8r\|A-B\|_{\mathrm{op}}^2
        \tag*{(by $\|A-\operatorname{proj}_r(A)\|_{\mathrm{op}}\le\|A-B\|_{\mathrm{op}}$)}
    \end{align*}
\end{proof}

\begin{lemma}[Projection onto quantum states]
\label{lem:state-projection}
    Let $A\in\mathbb C^{d\times d}$ be Hermitian with at least one positive eigenvalue. Let $A_+$ be obtained by replacing the negative eigenvalues of $A$ by zero, and define
    \[
        \Pi(A):=\frac{A_+}{\operatorname{Tr}(A_+)}.
    \]
    Then $\Pi(A)$ is a quantum state of rank at most $\mathrm{rank}(A)$. Moreover, for every quantum state $\rho$,
    \[
        \|\Pi(A)-\rho\|_{\mathrm{tr}}
        \le 2\|A-\rho\|_{\mathrm{tr}}.
    \]
\end{lemma}

\begin{proof}
    Positivity, unit trace, and the rank bound follow directly from the definition. Suppose first that $\operatorname{Tr}(A_+)>0$. Since $A_+\succeq 0$ and $A_+-A \succeq 0$ have orthogonal supports,
    \begin{align*}
        \|\Pi(A)-A\|_{\mathrm{tr}}
        &=
        \left\|
        \left(\frac{1}{\operatorname{Tr}(A_+)}-1\right)A_+ +(A_+-A)
        \right\|_{\mathrm{tr}}
        \\
        &=|1-\operatorname{Tr}(A_+)|
        +\operatorname{Tr}(A_+-A)\\
        &=
        \max\left\{
            1-\operatorname{Tr}(A),
            \|A\|_{\mathrm{tr}}-1
        \right\}
        \tag*{(by $\|A\|_{\mathrm{tr}}=2\operatorname{Tr}(A_+)-\operatorname{Tr}(A)$)}
        \\
        &=
        \max\left\{
            \operatorname{Tr}(\rho-A),
            \|A\|_{\mathrm{tr}}-\|\rho\|_{\mathrm{tr}}
        \right\}
        \tag*{(by $\operatorname{Tr}(\rho)=\|\rho\|_{\mathrm{tr}}=1$)}
        \\
        &\le \|A-\rho\|_{\mathrm{tr}}.
    \end{align*}
    Then the triangle inequality gives
    \[
        \|\Pi(A)-\rho\|_{\mathrm{tr}}
        \le
        \|\Pi(A)-A\|_{\mathrm{tr}}
        +\|A-\rho\|_{\mathrm{tr}}
        \le 2\|A-\rho\|_{\mathrm{tr}}.
    \]
\end{proof}

\subsection{Theoretical Guarantees}

\begin{theorem}[Low-rank shadow bootstrapping]
\label{thm:low-rank-bootstrapping}
    Let $\rho\in\mathbb C^{d\times d}$ be an $n$-qubit quantum state of rank at most $r$, where $d=2^n$. For every $0<\epsilon,\delta<1$, there is a nonadaptive single-copy shallow-depth protocol whose reconstruction outputs an estimate $\widehat\rho$ satisfying $\Pr[\|\widehat\rho-\rho\|_{\mathrm{tr}}\le\epsilon]\ge1-\delta$. The protocol uses $N$ measurement settings and $T$ copies, where
    \begin{equation*}
        N=\widetilde O(r),
        \qquad
        T=\widetilde O\!\left(\frac{dr^2}{\epsilon^2}\right).
    \end{equation*}
\end{theorem}
\begin{proof}
    Let $J:=\lceil\log_2(4r/\epsilon^2)\rceil$, and start from any fixed quantum state $\sigma_0$.
    
    For $t=1,\ldots,J$, choose
    \[
        N_t:=
        \left\lceil
            24C_0^2(C+1)r\log\frac{2dJ}{\delta}
        \right\rceil,
        \qquad
        S_t:=d\,2^{t-1},
    \]
    where $C_0$ is the constant in Lemma~\ref{lem:clipped-matrix-bernstein}. Let $\Delta_{t,i}$ be the residual estimator in Definition~\ref{def:bootstrap-residual}, with reference $\sigma_{t-1}$, using independent measurement data at each stage.
    
    The proof of Lemma~\ref{lem:bootstrap-residual-variance} also applies to Hermitian reference matrices $\sigma_{t-1}$ and gives
    \begin{align*}
        \mathbb E[\Delta_{t,i}\mid\sigma_{t-1}]
        &=\rho-\sigma_{t-1},\\
        \left\|
            \mathbb E[\Delta_{t,i}^2\mid\sigma_{t-1}]
        \right\|_{\mathrm{op}}
        &\le
        (C+1)\left(
            \|\rho-\sigma_{t-1}\|_F^2+\frac d{S_t}
        \right).
    \end{align*}
    Set
    \[
        K_t:=
        \sqrt{
            \frac{
                (C+1)
                \left(2/2^{t-1}+d/S_t\right)N_t
            }{\log(2dJ/\delta)}
        },
    \]
    and define
    \[
        \sigma_t
        :=
        \operatorname{proj}_r\!\left(
            \sigma_{t-1}
            +\frac1{N_t}\sum_{i=1}^{N_t}
            \operatorname{clip}_{K_t}(\Delta_{t,i})
        \right).
    \]

    Suppose that
    $\|\rho-\sigma_{t-1}\|_F^2\le2/2^{t-1}$.
    Applying Lemmas~\ref{lem:clipped-matrix-bernstein}
    and~\ref{lem:rank-r-truncation} gives
    \begin{align*}
        \|\sigma_t-\rho\|_F^2
        &\le
        8r\left\|
            \frac1{N_t}\sum_{i=1}^{N_t}
            \operatorname{clip}_{K_t}(\Delta_{t,i})
            -(\rho-\sigma_{t-1})
        \right\|_{\mathrm{op}}^2\\
        &\le
        \frac{8C_0^2(C+1)r\log(2dJ/\delta)}{N_t}
        \left(
            \frac{2}{2^{t-1}}+\frac d{S_t}
        \right)\\
        &\le \frac{2}{2^t},
    \end{align*}
    with conditional probability at least $1-\delta/J$.

    Since $\rho$ and $\sigma_0$ are quantum states, $\|\rho-\sigma_0\|_F^2\le2$. Induction and a union bound therefore show that these bounds hold at every stage with probability at least $1-\delta$. On this event, $\operatorname{rank}(\sigma_J-\rho)\le2r$ implies
    \[
        \|\rho-\sigma_J\|_{\mathrm{tr}}
        \le
        \sqrt{2r}\,\|\rho-\sigma_J\|_F
        \le
        \sqrt{\frac{4r}{2^J}}
        \le\epsilon.
    \]

    Finally, $2^J=O(r/\epsilon^2)$ gives
    \begin{align*}
        N
        &=\sum_{t=1}^{J}N_t
        =O\!\left(rJ\log\frac{2dJ}{\delta}\right)
        =\widetilde O(r),\\
        T
        &=\sum_{t=1}^{J}N_tS_t
        =O\!\left(
            rd\,2^J\log\frac{2dJ}{\delta}
        \right)
        =\widetilde O\!\left(\frac{dr^2}{\epsilon^2}\right).
    \end{align*}
    
    On the success event, $\|\sigma_J-\rho\|_{\mathrm{tr}}\le\epsilon$ ensures that $\sigma_J$ has at least one positive eigenvalue.
    By Lemma~\ref{lem:state-projection}, replacing $\epsilon$ by $\epsilon/2$ and projecting the final estimate onto the set of quantum states gives a positive semidefinite, unit-trace output with error at most $\epsilon$, without changing the asymptotic bounds on $N$ and $T$. All measurement bases and shot counts are chosen in advance, and each measurement circuit has depth $O(\log n)$ by Lemma~\ref{appx:lem:suffix-ensembles}.
\end{proof}
\end{document}